%% file: main.tex
\documentclass[longauth]{aa}

\usepackage{graphicx}
\usepackage{natbib}
\usepackage{scalerel}
\usepackage{booktabs}

\usepackage[table]{xcolor}

\usepackage{txfonts}
\usepackage[pdfencoding=auto,psdextra]{hyperref}
\hypersetup{
    colorlinks=true,
    linkcolor=blue,
    filecolor=magenta,      
    urlcolor=blue,
    citecolor=blue
}
\usepackage{cleveref}
\crefname{figure}{Fig.}{Fig.}
\crefname{table}{Table}{Table}
\crefname{equation}{Eq.}{Eq.}
\crefname{section}{Sect.}{Sect.}

\crefformat{equation}{Eq.~(#2#1#3)}
\Crefformat{equation}{Eq.~#2#1#3}

\makeatletter
\renewcommand*\aa@pageof{, page \thepage{} of \pageref*{LastPage}}
\makeatother

\usepackage[utf8]{inputenc}

\usepackage{euclid}

\newcommand{\ADS}{$A_\mathrm{DS}$}
\newcommand{\sqrtADS}{$\sqrt{A_\mathrm{DS}}$}
\newcommand{\Gaia}{{\it Gaia\/}}
\newcommand{\zstar}{$\Delta z_\mathrm{star}$}
\newcommand{\zobs}{$\Delta z_\mathrm{obs}$}

\begin{document}
%
% Put the title and authors of your (Standard Project) paper here
%

% \title{\Euclid\/: Precise inference of defocus wavefront error from image diffraction spike measurements\thanks{This paper is published on
%        behalf of the Euclid Consortium}}

\title{\Euclid: Precise inference of defocus wavefront error from image diffraction spike measurements\thanks{This paper is published on behalf of the Euclid Consortium.}}    

%% please do not edit the author list once you copy it from the
%% Publication Portal -- contact ECEB Bureau for changes
   
% \newcommand{\orcid}[1]{} %% define as link to https://orcid.org/#1 if needed
% \author{\normalsize D.~Neumann$^{1}$\thanks{\email{dneumann@strw.leidenuniv.nl}}, L.~Miller$^{2}$, H.~Hoekstra$^{1}$, K.~Kuijken$^{1}$, I.~Whittam$^{2}$, N.~E.~Chisari$^{3,1}$ \emph{et al}}
\include{authors}

%
% For a Key Project paperm please use instead:
%
% \title{\Euclid\/ preparation. TBD. Your title}
%
% \author{Euclid Collaboration: F.~Author, .....}
%

% 
% Put your abstract here:
%
\abstract{
The success of the \Euclid cosmological weak lensing measurements requires unprecedented knowledge of the point spread function (PSF) shape.
Defocus wavefront errors induce variations in PSF size that directly bias inferred galaxy shapes.
We present a rapid, model-independent estimator of the \Euclid visible instrument defocus based on measuring subpixel shifts in diffraction spikes from bright stars, arising from the non-mirror-symmetric placement of the telescope spiders.  
This estimate requires no a-priori PSF model and can be directly inferred from individual survey exposures within seconds.
We express defocus as the secondary mirror displacement along the optical axis, $\Delta z$, achieving a per-exposure precision of $\sigma(\Delta z) = 0.022\,\micron$ in standard \Euclid wide-survey images, corresponding to a peak-to-valley optical-path difference of $\qty{0.75}{\nano\metre}$.
This sensitivity resolves thermally induced shifts, typically 0.1--0.3$\,\micron$, and enables tracking of temporal evolution and field-of-view variations when combining exposures within periods of stability.
Since July 2024, the \Euclid telescope has been exceptionally stable, while changes to defocus are effectively homogeneous across the field of view. 
By correlating defocus with PSF size, we find that the \Euclid DR3 requirement of a fractional PSF size bias of $\left|\Delta R_\mathrm{PSF}^2/R_\mathrm{PSF}^2\right|<10^{-3}$ corresponds to field-averaged secondary mirror displacements exceeding $\langle \Delta z_\mathrm{thr}\rangle = (0.0683 \pm 0.0024)\,\micron$. Our estimator provides a robust real-time monitoring tool, supplies a stringent prior for computationally intensive PSF model fitting, and is applicable not only to the \Euclid telescope, but to other telescopes whose spider vanes are not mirror-symmetrically arranged.
}
%
% Provide up to five key words:
%
    \keywords{Cosmology: observations -- Gravitational lensing: weak -- Instrumentation: detectors -- Techniques: image processing}
%    from the list in
%     https://www.aanda.org/for-authors/latex-issues/information-files#pop}
%
% Add short versions of title and author list for page headings
%
\titlerunning{\Euclid: Inferring defocus from diffraction spikes}
\authorrunning{Neumann et al.}
\maketitle
\nolinenumbers
%
%-------------------------------------------------------------------
%
%
%   Start the main text of your paper here
%
   
\section{\label{sc:Intro}Introduction}

One of the primary probes of the \Euclid surveys is weak gravitational lensing \citep[see][]{hoekstra_weak_2008, mandelbaum_weak_2018, PRAT2026508}, which traces the coherent distortion of galaxy light by deflection along the gravitational potential caused by the large-scale structure (LSS) of the Universe. 
Over its nominal six-year survey, the \Euclid telescope will measure the shapes of 1.5 billion galaxies spanning roughly a third of the sky, allowing weak gravitational lensing analyses to be performed with unprecedented statistical precision \citep{EuclidSkyOverview}.

This large statistical power will allow tight constraints on the evolution of cosmic structure over billions of years. However, the accuracy of these constraints ultimately depends on the control of systematic uncertainties.
For galaxy shape measurements, one of the main sources of systematic error is the point spread function (PSF), which describes the image of a point source distorted by the telescope's optical system \citep{kaiser_method_1995, bartelmann_weak_2001, bernstein_shapes_2002}. 
The PSF convolution alters the observed shapes of galaxies, and errors in its modelling propagate directly into biases in cosmological inference \citep{hirata_shear_2003, huterer_systematic_2006, paulin-henriksson_point_2008, cropper_defining_2013}.
Although \Euclid, operating at the second Lagrange point (\Ltwo), achieves one of the smallest and most stable PSFs of any cosmological survey to date, its spatial and temporal variations must still be known to the 0.1\% level regarding size and 0.01\% level regarding ellipticity to disentangle instrumental shape correlations from true gravitational lensing signals \citep{Laureijs11,massey_origins_2013}.
To meet these requirements, the Euclid Consortium employs forward modelling of the visible wavelength (VIS) instrument’s optical PSF, including field-of-view (FoV) dependent wavefront error (WFE) terms (Euclid Collaboration: Miller et al. 2026, in prep.).
% \citep{DR1-TP018} -> In prep to be exchanged with this for DR1

Among all WFE components in the \Euclid payload module, defocus is expected to be the most variable, primarily driven by thermal fluctuations (\citealt{E-Anselmi}; Euclid Collaboration: Whittam et al. 2026, in prep.), in line with previous experience with the \HST \citep{krist_20_2011}.
% \citep{E-Anselmi, DR1-TP021} -> exchange reference above with DR1
Continuous information about defocus is therefore valuable both for monitoring the optical state of the spacecraft and for informing or constraining the PSF forward-modelling pipeline.
%Past experiences with the Hubble Space Telescope (HST)  revealed correlations between temperature fluctuations and defocus-WFE \citep{krist_20_2011} and highlighted the necessity of regularly updating the model based on WFE calibrations. Thus, understanding the evolution of the defocus on an exposure-to-exposure basis will enable us not only to model the PSF more accurately and remove biases, but also provide a gateway to monitor the spacecraft itself.

Diffraction spikes are common artefacts in astronomical telescopes caused by diffraction at the secondary mirror (M2) spider support struts, in front of the entrance pupil.
Although most studies have focused on mitigating the artefacts these spikes introduce \citep{harvey_diffraction_1995, pueyo_highcontrast_2013, harvey_novel_2018}, their geometric properties encode information about the telescope's underlying WFE.
This idea is conceptually similar to the use of a pupil mask, such as Bahtinov masks for focusing small telescopes \citep{Bahtinov2005}.
However, here it is applied directly to the diffraction patterns that naturally form around bright stars in \Euclid images taken with the visible imaging instrument (VIS). Therefore, this method requires no additional optics or wavefront sensors.

In this work, we develop a diffraction spike-based method to estimate the VIS defocus directly from imaging data and apply it to \Euclid data. 
Without prior PSF model information, we measure the FoV-dependent defocus and assess its temporal stability, with particular emphasis on the \Euclid weak lensing cosmology analysis requirements.
To our knowledge, this is the first study to employ the entrance pupil struts of a telescope as direct wavefront sensors.

\begin{figure*} 
    \centering
    \includegraphics[width=0.85\linewidth]{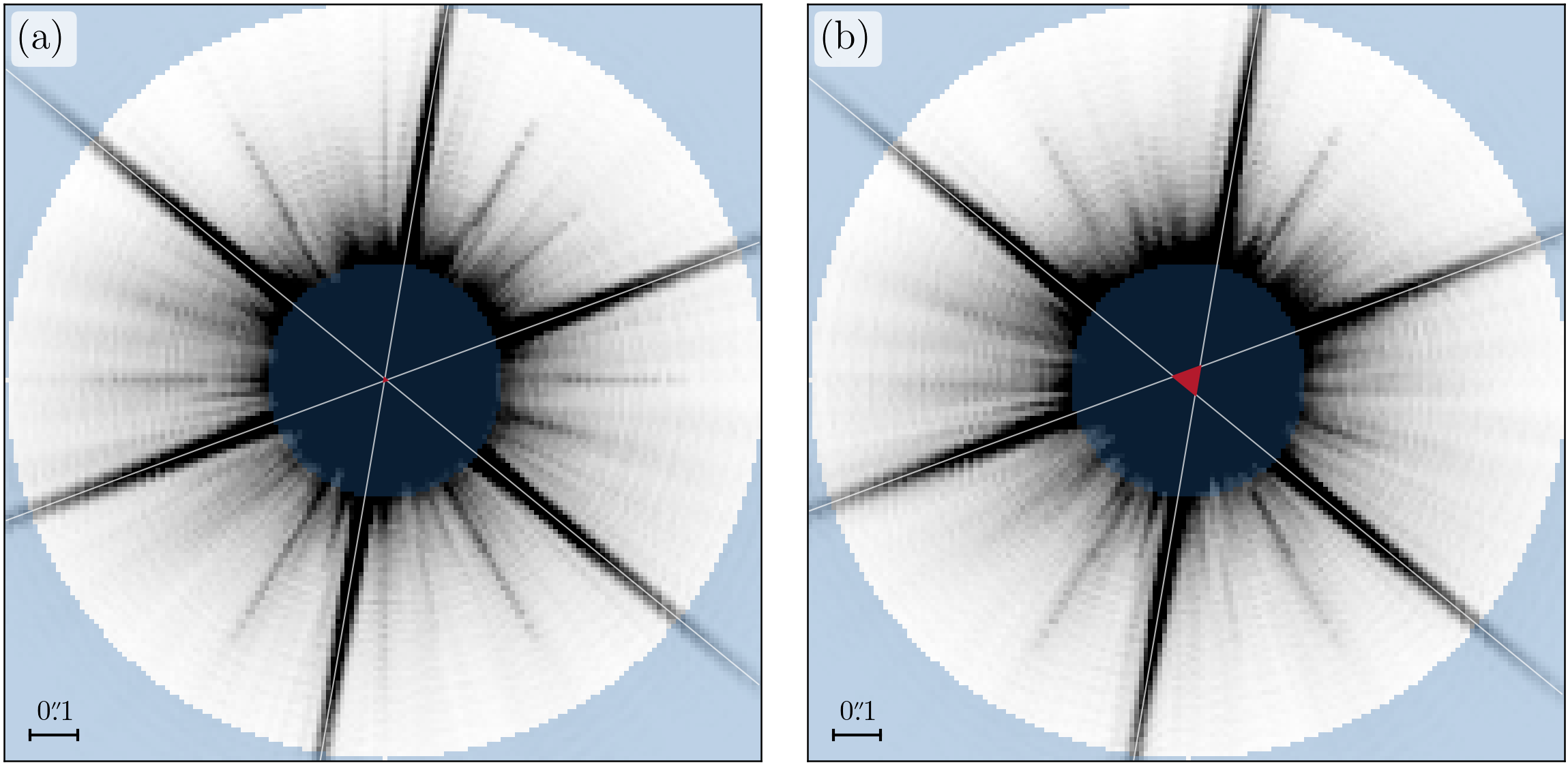}
    \caption{Simulated \Euclid PSF nearly in-focus (a) and defocused by 12.25\,\micron{} (b). The detected diffraction spike positions are indicated with grey lines and their enclosed triangle is coloured red. Blue shaded regions are defined by $r<\ang{;;2.5}$ and $r>\ang{;;8.0}$ and are masked during the spike detection.}
    \label{fig:defocus_example}
\end{figure*}

We will first motivate the relation between defocus WFE and the shift in diffraction spike position in \cref{sc:Background} and its effect on the size of the \Euclid PSF model.
Thereafter, we outline our methodology in \cref{sc:Methodology}, starting with our diffraction spike detection method. 
We further explain how the shift in spikes can be translated into the physically interpretable shift in optical elements (defocus), estimate the statistical error per single star (\cref{sc:Methodology-Expected Uncertainty}), and outline how we estimate the defocus value per \Euclid exposure (\cref{sc:Methodology-Iterative defocus calculation}).
In \cref{sc:Results}, we investigate the temporal evolution of the defocus by examining the change of its FoV dependence over time, demonstrating the remarkable stability of the spacecraft. 
We show that this method can be used as a strong prior for PSF modelling. 
We summarise our findings in \cref{sc:Conclusion}.

\section{\label{sc:Background}Background}

The \Euclid mission and its optical VIS instrument are described in \citet{EuclidSkyOverview}, \citet{EuclidSkyVIS}, and \citet{Q1-TP002}. The PSF model is described in Euclid Collaboration: Miller et al. (2026, in prep.). In the following sections, we will focus on the information we can extract from the diffraction spikes.
% \citet{DR1-TP018} -> exchange for in prep.

\subsection{\label{sc:Background-Relation between diffraction spike and defocus}Relation between diffraction spike position and defocus}

In general, the WFE $W_\mathrm{E}(r,\phi)$ describes the optical-path difference between an ideal wavefront and the actual wavefront projected onto the unit disc in polar coordinates $(r,\,\phi)$.
A uniform shift of the image plane along the optical axis, $\vec{\hat{z}}$, leads to a defocus of the image. 
The resulting isotropic wavefront error can be described by 
\begin{align}
    \label{eq:defocus_WFE}
    W_\mathrm{E}(r) &= -\frac{r^2}{2}\left(\frac{1}{f+\Delta z_\mathrm{eff}}-\frac{1}{f}\right) \overset{\Delta z_\mathrm{eff}\ll f}{\approx}  d\,\left(\frac{2r}{D}\right)^2
\end{align}
with
\begin{align}
    d &\coloneqq \frac{\Delta z_\mathrm{eff}}{8\,N^2}\,,
\end{align}
where $d$ is the peak-to-valley optical-path difference, $r$ the distance to the optical axis centre, $f$ the focal length, and $\Delta z_\mathrm{eff}$ the shift of the image plane along the optical axis \citep[Chapter 6.4.4]{goodman_introduction_1996}.
Furthermore, $N=f/D$ is the optical system's $f$-number and $D$ the effective entrance pupil diameter; in the case of the \Euclid VIS instrument, $N=20.4167$.
The shift of the focal plane $\Delta z_\mathrm{eff}$ may be related to a shift of the secondary mirror along the optical axis $\Delta z$ by the linear relation
\begin{align}
    \label{eq:z_eff}
    \Delta z_\mathrm{eff} = \alpha\,\Delta z\;,
\end{align}
where $\alpha$ depends on the telescope design, with $\alpha\simeq113.8$ in the case of \Euclid, as obtained from an Airbus Defence and Space \texttt{CodeV}\footnote{\url{https://www.keysight.com/us/en/products/software/optical-solutions-software/optical-design-solutions/codev.html}} model of the as-built Euclid telescope (Pierre-Antoine Frugier, private communication).
In \Cref{apdx:A}, we show that such a wavefront error causes a lateral shift of the diffraction spikes $\Delta x$ that is directly proportional to the shift of the image plane, i.e. $\Delta x\propto\Delta z$. 
In \cref{fig:defocus_example} this is illustrated with PSF simulations obtained from the PSF model included in the \Euclid shear pipeline (\citealt{EuclidSkyOverview}; Euclid Collaboration: Miller et al. 2026, in prep.): (a) shows the PSF close to in-focus position and (b) a strongly defocused PSF, where the detected spike positions are shown with grey lines. 
% \citep{EuclidSkyOverview, DR1-TP018} -> exchange citation above what is the difference between aa and aaEC.sty?with this for DR1
Each spike shifts coherently inward or outward with defocus, depending on the direction of $\Delta z$.
As such, the enclosed area will form a triangle (approximately equilateral), where the shortest distance between each line and the centre is $\Delta x$.
One can easily show that the area of such a triangle is given by $A_\mathrm{DS}=3\sqrt3\,(\Delta x)^2$, and it directly follows that
\begin{equation}
    \label{eq:diffspike_propto_deltaz}
    \sqrt{A_\mathrm{DS}}\propto\Delta x\propto\Delta z\;.
\end{equation}
Thus, we adopt the square root of the enclosed diffraction spike triangle area, $A_\mathrm{DS}$, as the metric of interest in this work.
This triangle area increase with defocus is clearly visible in \cref{fig:defocus_example}, indicated with the red shaded region.
The size of the triangle is a direct measurement of defocus, while its orientation indicates the sign of $\Delta z$ for symmetry reasons. 
Thus, we assign a sign to $\sqrt{A_\mathrm{DS}}$ based on the orientation of the triangle, thereby encoding the direction of the defocus. 
In doing so, we effectively average the WFE sampled by the three struts supporting M2, which suppresses anisotropic contributions.
\citet{E-Anselmi} show that, during nominal survey operation, the defocus mode dominates the WFE variability by approximately an order of magnitude. Consequently, to first order, the diffraction spikes trace the full WFE fluctuations.

Nevertheless, in \Cref{apdx:A} we briefly examine the impact of higher-order WFE terms and demonstrate, using a physical PSF model (Euclid Collaboration: Miller et al. 2026, in prep.), that our estimator is most sensitive to the defocus Zernike mode (ANSI index 4), with roughly half that sensitivity to vertical trefoil (ANSI index 6). 
% \citep{DR1-TP018} -> exchange above
We further show in Euclid Collaboration: Whittam et al. (2026, in prep.) that the trefoil contribution to our $\Delta z$ estimator is small following the June 2024 ice-decontamination campaign discussed in \cref{sc:Results-deicing-influence}. Moreover, Euclid Collaboration: Whittam et al. (2026, in prep.) present a methodology to remove the residual sensitivity to trefoil.
% \citet{DR1-TP021} -> exchange both above

It must be noted that, fortuitously, the \Euclid spacecraft is uniquely well designed for defocus inference with diffraction spikes: The 560.52\,s long exposures in combination with the large FoV yield typically $\mathcal{O}(100)$ bright stars per exposure with clearly visible spikes.
Additionally, the non-mirror symmetry of the individual \Euclid telescope spider arms produces non-overlapping diffraction spikes, allowing us to determine both the magnitude and direction of defocus.

Similarly, the \HST quantifies its defocus change in terms of secondary mirror displacement as well.
\emph{Hubble} shrank by $\sim$$150\,\micron$ since launch due to water desorption from its graphite trusses \citep{krist_20_2011}, while its focus varies by 3--10$\,\micron$ because of its constantly changing solar aspect angle \citep{rhodes_stability_2007,cox_2011_evaluation}. Finally, it `breathes' by $\sim$$2\,\micron$ every few hours when it passes in and out of the shadow of the Earth \citep{DiNino_2008_HST}.  

\subsection{\label{sc:Background-Relation between defocus and PSF size}Relation between defocus and PSF size}

In addition to a shift in diffraction spikes, a defocus will naturally affect the size of the PSF \citep{stokseth_properties_1969}. 
In fact, the PSF diameter is most sensitive to the defocus WFE (Whittam et al. 2026, in prep.) and grows proportional to $\Delta z$ for small defoci \citep{subbarao_1994_depth}\footnote{\citet{subbarao_1994_depth} use geometric optics which is known to be a mediocre approximation for diffraction limited systems \citep{stokseth_properties_1969}. In our work, we verified using \texttt{GalSim} \citep{rowe_galsim_2015} that $R^2=Q_{11}+Q_{22}$ defined by \cref{eq:quadrupole_brightness_moments} grows with $(\Delta z)^2$ within per cent accuracy for a \Euclid like telescope within $\pm0.1$ wavelength of pure defocus, using $\lambda_\mathrm{fid}=700\,\mathrm{nm}$.}, and by extension, to \sqrtADS{}.
% \citet{DR1-TP021} -> exchange above
In this work, we quantify the PSF size with brightness moments, $Q_{ij}$, that are also commonly employed in galaxy shape measurement pipelines \citep{kaiser_method_1995, hoekstra_weak_1998, melchior_weak_2011}.
In accordance with the \emph{Euclid} requirements \citep{paulin-henriksson_point_2008, Laureijs11}, we adopt $R^2=Q_{11}+Q_{22}$ as the measure of PSF size, where $Q_{11},\,Q_{22}$ are quadrupole brightness moments defined by
\begin{align}
    \label{eq:quadrupole_brightness_moments}
    Q_{ij}= \iint\diff^2 r\;W(\boldsymbol{r})\,I(\boldsymbol{r})\,r_i\,r_j\,,
\end{align}
where $\boldsymbol{r}=(x,y)$ is the two component image position vector with origin at the location where the dipole vanishes, $I(\boldsymbol{r})$ the pixel brightness and $W(\boldsymbol{r})$ a weight function to suppress background noise that would otherwise dominate the integral for large $\boldsymbol{x}$.
Note that the integral has to be evaluated over an image stamp large enough not to clip the weight function.
For shape measurements, $W(\boldsymbol{r})$ is commonly defined as a radially symmetric Gaussian distribution.
This PSF size parametrisation is unconventional in an optics setting but has an advantage in this context:
the modelling error of this size metric defined as 
\begin{align}
    \label{eq: DeltaR2/R2}
    \Delta R_\mathrm{PSF}^2/R_\mathrm{PSF}^2 \coloneqq \frac{R^2_\mathrm{model}-R^2_\mathrm{true}}{R^2_\mathrm{true}}\,,
\end{align}
propagates directly into the systematic uncertainty on the cosmological weak lensing inference \citep{paulin-henriksson_point_2008,amara_systematic_2008, massey_origins_2013}, motivating the requirement $\left|\Delta R_\mathrm{PSF}^2/R_\mathrm{PSF}^2\right|<10^{-3}$ on the \Euclid PSF model \citep{cropper_defining_2013}.
In \cref{sc:Results-Correlation of Defocus with PSF core size} we investigate how this relates to the actual change in our defocus estimate over nominal survey operations. 

\section{\label{sc:Methodology}Methodology}

\subsection{\label{sc:Methodology-Spike detection}Spike detection method}

The methodology presented here requires no a-priori model information and can thus be easily generalised to other telescope configurations, given obstructive straight, thin elements in the entrance pupil that do not exhibit mirror symmetry (see \Cref{apdx:pupil_configurations}).
For symmetric elements, defocus still induces an apparent shift of the PSF diffraction spikes.
However, the actual shift will be obscured by the fact that multiple spikes overlap for small defoci, as visualised in \Cref{apdx:pupil_configurations}.
We demand that first-order brightness moments vanish at the spike centre $(x_\mathrm{c},\;y_\mathrm{c})$, such that
\begin{equation}
    \label{eq:centering_with_brightness_moments}
    \iint\mathrm{d}x\;\mathrm{d}y\;W(x-x_\mathrm{c},\;y-y_\mathrm{c})\;
    \begin{pmatrix}
        x-x_\mathrm{c} \\
        y-y_\mathrm{c}
    \end{pmatrix}
    \;I(x,y) = \vec{0}\,,
\end{equation}
where $I(x,y)$ is the pixel brightness and $W$ an appropriate weight function. 
Note that $W(\boldsymbol{x})$ in this case will be significantly different to the one commonly used for galaxy shape measurements.
The procedure to extract the area $A_\mathrm{DS}$ (see \Cref{eq:diffspike_propto_deltaz}) is then as follows:
\begin{itemize}
    \item We iteratively apply \cref{eq:centering_with_brightness_moments} with a Gaussian weight function $W=\mathcal{N}(x-x_\mathrm{c},y-y_\mathrm{c},\sigma)$ for $\sigma=1\arcsecond$ until the position where the dipole vanishes is known to a precision of $0.01\,$pixel.
    We equate this with the centre of the star.
    This is relevant only for the subsequent masking, and pixel level inaccuracies will have negligible effect thanks to the masking apertures.
    \item The inner and outer part of the star are masked with a circular aperture of $r_\mathrm{inner}=25\,$pixel and $r_\mathrm{outer}=80\,$pixel, visually chosen to conservatively mask out PSF core light even for the brightest stars we consider (inner bound) and to remove noise due to the decaying diffraction spike brightness (outer bound), also shown in \cref{fig:defocus_example}. 
    Note that \cref{fig:defocus_example} contains no noise; hence we exaggerate the visibility of the PSF wings.
    \item On both opposing sides ($+/-$) for each spike, we again iteratively apply \cref{eq:centering_with_brightness_moments} with a Gaussian tophat function as weight defined as
    \begin{equation}
        W_\pm(x-x_\mathrm{c},\,y-y_\mathrm{c},\,\sigma,\,h)= \exp{\left(-\frac{(x')^2}{2\sigma^2}\right)}\,\Theta(h/2\mp|y'|)\,,
    \end{equation}
    where $\Theta(x)$ is the Heaviside step function, $h$ the longitudinal size of the weight along the diffraction spike, chosen here as $h=46\,$pixel, and $\sigma=2\,$pixel the width of the Gaussian in the transverse direction.
    Note that we leave a 5\,pixel padding region between $r_1$ and $W_{\pm}$ to avoid masked pixels.
    We use rotated coordinates ($x',\,y'$), such that $y'$ is parallel to the spike direction and $x'$ perpendicular.
    With a given known angle $\theta$ they can be written as
    \begin{align}
        \begin{pmatrix}
        x' \\[4pt]
        y'
        \end{pmatrix}
        =
        \begin{pmatrix}
        \cos\theta & \sin\theta \\
        -\sin\theta & \cos\theta
        \end{pmatrix}
        \begin{pmatrix}
        x - x_{\mathrm{c}} \\[4pt]
        y - y_{\mathrm{c}}
        \end{pmatrix}\;.
    \end{align}
    In each iteration, we project the inferred shift of the centre ($x_\mathrm{c},\, y_\mathrm{c}$) to the ${x}'$ direction to prevent the centre from drifting along the $y'$ direction. A typical shift is small, on the order of fractions of a pixel.
    We obtain a central spike position from either side of the star for all three visible spikes. We found empirically that the angles of $\theta \in \{\ang{170.44},\,\ang{51.02},\,\ang{110.13}\}$ with respect to the focal plane array (FPA) vertical are a suitable starting point for robust detections.
    \item We then convert the two opposite central positions per spike into line equations from which we obtain the three intersection points $(x_1,\,y_1),\,(x_2,\,y_2),\,(x_3,\,y_3)$. 
    The enclosed area is then easily obtained using 
    \begin{align}
        \label{eq:diffspike_area_from_intersection}
        A_\mathrm{DS} = \frac{1}{2} \left[ x_{1}(y_{2} - y_{3}) + x_{2}(y_{3} - y_{1}) + x_{3}(y_{1} - y_{2}) \right]\,.
    \end{align}
    Note that we allow $A_\mathrm{DS}$ to be negative to preserve information about the triangle orientation, making \cref{eq:diffspike_area_from_intersection} different to its usual form \citep{anton_elementary_2010}. This enables us to break degeneracy of the defocus direction by modifying \cref{eq:diffspike_propto_deltaz} to $\sqrt{|A_\mathrm{DS}|}\,\mathrm{sgn}(A_\mathrm{DS})\propto \Delta z$. For clarity, we will still refer to this as $\sqrt{A_\mathrm{DS}}$ for the remainder of this work, aware that this can be a negative quantity due to the direction of defocus.
\end{itemize}
The procedure above takes $\sim$\,50 milliseconds per star using a single thread on an \texttt{AMD EPYC 9454} CPU that has a clock frequency of up to 3.8\,GHz.

\begin{figure}[t]
    \centering
    \includegraphics[width=0.9\linewidth]{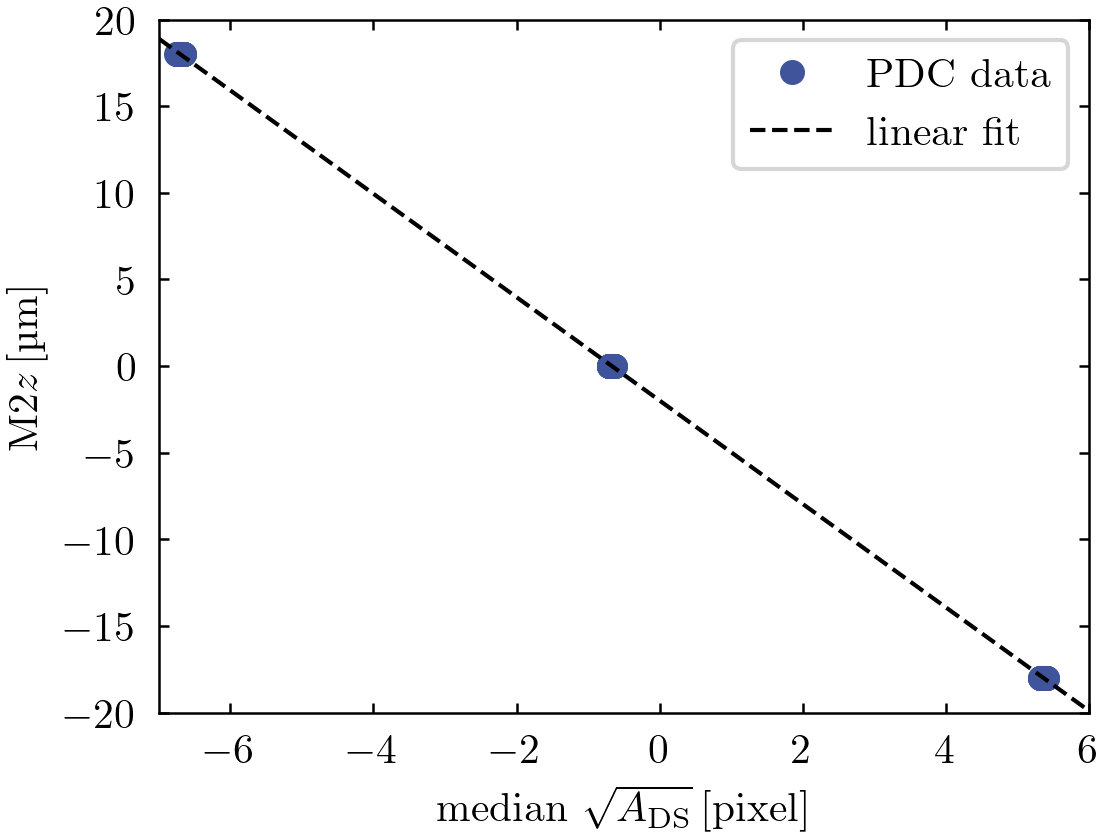}
    \caption{Relation between enclosed diffraction spike triangle area \ADS{} and actual mirror defocus M2$z$. Blue points are obtained by taking the median of $\sqrt{A_\mathrm{DS}}$ of all stars measured in a single exposure. The black dashed line is a linear fit the about 150 exposures per M2$z$-position taken as part of the PDC.}
    \label{fig:sqrtA_to_m2z}
\end{figure}

\subsection{\label{sc:Methodology-Calibration}Calibration}

To relate $A_\mathrm{DS}$ to a physically interpretable quantity, we utilise the phase diversity calibration (PDC) measurements of the \Euclid telescope (\citealt{laureijs_euclid_2024}; Euclid Collaboration: Duncan et al. 2026, in prep.) to find a conversion between \ADS{} and defocus in terms of secondary mirror shift $\Delta z$. 
% \citep{laureijs_euclid_2024, DR1-TP019} -> exchange above
The PDC data taken in early 2024 consist of $289$\,s exposures observed over 24 hours in each of three different $z$ positions of M2 along the optical axis.
The purpose of the PDC is to investigate the variation of WFE caused by changes in the telescope state, obtained by intentionally shifting M2 while the telescope is thermally stable. 
In this work, we use the PDC observations taken between 2024/01/09 and 2024/01/11, amounting to 150 exposures of the same pointing in extra-focal position (M2$z\simeq+18\,\micron$), intra-focal position (M2$z\simeq-18\,\micron$) and in-focus position (M2$z\simeq0\,\micron$), respectively.\footnote{In practice, due to a download issue, we only use 149 exposures for the extra-focal and in-focus case, amounting to a total of 448 exposures here.}
The bright stars we consider in this work have been measured by the \Gaia{} survey \citep{prusti_gaia_2016, vallenari_gaia_2023} which we crossmatch with the \Euclid exposures for auxiliary information.
Note that we do not explicitly filter binary stars since the majority of them are unresolved.
The fraction of resolved binaries, here defined as stars with separation $<\ang{;;0.1}$ and different \emph{Gaia} IDs, comprises $\lesssim$0.25\% of the total sample; a negligible fraction.
With the procedure described in \cref{sc:Methodology-Spike detection}, we obtain the \ADS{} values for the $\sim\,$160 stars between \Gaia{} $G$ magnitude $11.5<m_G<15.5$ for each exposure. 
The reason for this exact cutoff is detailed further in \cref{sc:Methodology-Expected Uncertainty}. 
We then take the median value of $\sqrt{A_\mathrm{DS}}$ and fit a linear relation between M2$z$ and $\sqrt{A_\mathrm{DS}}$ based on the three known defocus positions. The fit is shown in \cref{fig:sqrtA_to_m2z} as a black dashed line. The 448 median \sqrtADS{} show a remarkably small spread in each respective M2$z$ position. The obtained relation is
\begin{align}
    \label{eq:diffspike_to_defocus}
    \mathrm{M2}z =& -2.98352(60)\,\frac{\micron}{\mathrm{pixel}}\,\sqrt{A_\mathrm{DS}} - 1.9912(30)\,\micron\,.
\end{align}
\begin{figure}[t]
    \centering
    \includegraphics[width=1\linewidth]{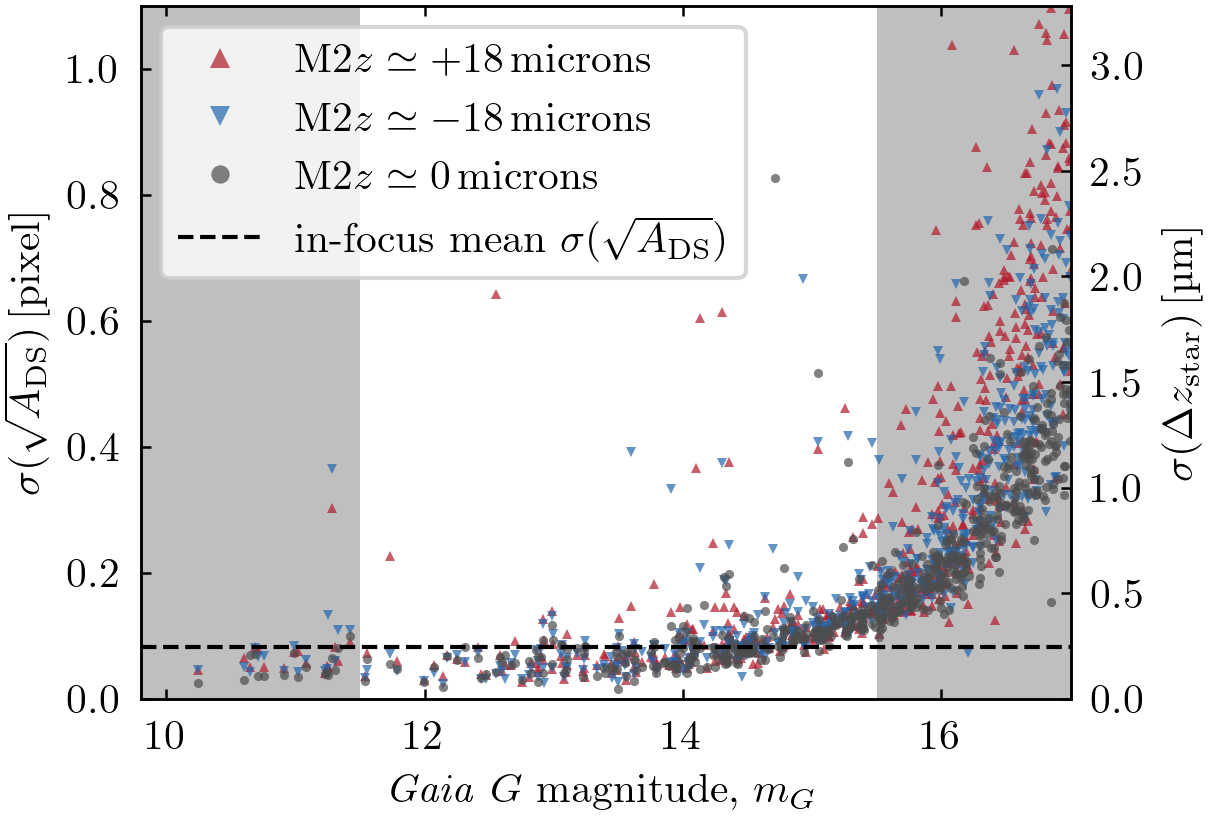}
    \caption{Standard deviation of \sqrtADS{} (left axis) and defocus (right axis) as a function of \Gaia{} $G$ magnitude for repeated measurements of individual stars. Red, blue, and grey points respectively indicate extra-focal ($+$18\,\micron), intra-focal ($-$18\,\micron), and in-focus (0\,\micron) M2 position along the optical axis. Grey bands indicate magnitudes that are disregarded for the calibration. The black line shows the mean $\sigma\left(\!{\sqrt{A_\mathrm{DS}}}\right)$ of the in-focus measurements.
    The data were obtained during the \Euclid phase diversity calibration campaign.}
    \label{fig:std_over_Gmag}
\end{figure}
By taking the median, we implicitly ignore any FoV-dependent behaviour that is generally present. 
In fact, the FoV dependency is one of the main points of emphasis in this work. 
However, we empirically show in \Cref{apdx:non_uniformity_of_defocus_change} that the defocus changes homogeneously across the FoV for the major part of the Euclid Survey, allowing us to use the slope of \cref{eq:diffspike_to_defocus} for individual stars, as opposed to the median only.
Additionally, the slope between M2$z$ and \sqrtADS{} for individual quadrants varies only at the $1$\% level.
We will henceforth differentiate between the defocus of individual stars, denoted as \zstar{}, and the summarised defocus of an exposure, denoted as \zobs{} (corresponding to $\Delta z$ from \Cref{eq:z_eff}).
As we show later, the variation of \zstar{} across the FoV represents the general WFE structure without additional distortions. 
We refer to this as the `baseline' FoV dependency, or the baseline defocus FoV distribution throughout this work.
In practice, \zstar{} represents a convenient scaling of \sqrtADS{} that allows for direct magnitude comparisons between \sqrtADS{} and \zobs{}.
The defocus \zobs{}, on the other hand, is the relative change of \zstar{} per exposure, absorbed into a single number.
Note that we effectively equate M2$z$ and \zobs{}, although \zobs{} is thought to originate from thermo-mechanical breathing of the spacecraft optical structure (\citealt{E-Anselmi}; Euclid Collaboration: Whittam et al. 2026, in prep.).
However, both produce a homogeneous change in \zstar{}, indicating similar effects on the optical path.

\subsection{\label{sc:Methodology-Expected Uncertainty}Estimated uncertainty in defocus per individual star and per \Euclid exposure}

The individual PDC exposures have all the same pointing, allowing us to evaluate the precision of our inferred defocus.
The main changes between exposures are different noise realisations and slight pointing variations, thus allowing us to assess the systematic uncertainty of our method by looking at the standard deviation of \sqrtADS{} for individual stars, giving us a direct measure of overall statistical uncertainty.
We expect the signal-to-noise ratio of the spikes to be the main source of variability here. 

In \cref{fig:std_over_Gmag} we show the resulting standard deviation per star in terms of \sqrtADS{} (left $y$-axis) and \zstar{} (right $y$-axis) as a function of the \Gaia{} \emph{G} magnitude ($m_G$) of the star. 
Red, blue, and grey points indicate single star standard deviations for the extra-focal, intra-focal, and in-focus position, respectively. 
The grey regions indicate excluded magnitudes. 
Extremely bright stars have profound vertical stripes due to CCD-bleeding. 
The bright streaks become visible for $m_G<13.5$ but are noticeably interfering with the spike detection algorithm for $m_G<10$, hence we choose a conservative lower cutoff at $m_G<11.5$.
We determine the upper cutoff of $m_G<15.5$ via the rapidly increasing slope in standard deviation discernible in \cref{fig:std_over_Gmag}. 
The majority of outliers are caused by artefacts in the images, such as ghosts or improperly masked cosmic rays that cross the diffraction spikes.
In general, strong outliers occur also due to pixel pollution by CCD artefacts or neighbouring sources, although few in number, rendering the use of the mean as summary statistic inadequate.

For the majority of stars however, the methodology described in \cref{sc:Methodology-Spike detection} reliably finds the centre of all spikes. 
In the PDC measurement, 4.86\% (2.02\%) of standard deviations in the in-focus sample lie 1.5 (3) times the interquartile range above the 75th percentile and can thus be interpreted as (strong) outliers.
Notably, this is just an indicator for stability and does not say anything about potential biases. These will be investigated in \cref{sc:Results-star-defocus-properties}.

We emphasise that we determine \sqrtADS{} up to fractions of a pixel, which in turn corresponds to fractions of \micron{} in the actual estimated defocus, measured in terms of secondary mirror shift. 
The median standard deviation of estimated defocus for stars with $11.5<m_G<15.5$ in the in-focus PDC exposures is 0.247\,\micron.
The nominal survey observations contain a median of 126 stars per exposure in the equivalent magnitude range (see Eqs. \labelcref{eq:g_mag_lower_bound} and \labelcref{eq:g_mag_upper_bound}), resulting in an approximate standard error in the mean of order 0.022\,\micron.
With \cref{eq:defocus_WFE}, we can convert this to a precision of \qty{0.75}{\nano\metre} for the actual peak-to-valley optical-path difference.
Therefore, we expect to be able to constrain the defocus of the \Euclid space telescope to the sub-percent level of a single wavelength per nominal science exposure. 

Nominal \Euclid exposures have an exposure time of $560.52$\,s. 
Therefore, the magnitude limits from the PDC observations need to be adjusted accordingly.
The faint limit is caused by the signal-to-noise ratio of the spikes that grows with $\sqrt{t}$ in the background-limited case. 
The bright limit originates from over-saturation that depends on the signal counts, thus growing $\propto t$.
This yields the limits
\begin{align}
    \label{eq:g_mag_lower_bound}
    m_{G,\mathrm{min}}^\mathrm{nominal} &\simeq12.22\,, \\
    \label{eq:g_mag_upper_bound}
    m_{G,\mathrm{max}}^\mathrm{nominal} &\simeq15.86\,.
\end{align}

\subsection{\label{sc:Methodology-Iterative defocus calculation}Iterative defocus calculation}

\begin{figure}[t]
    \centering
    \includegraphics[width=1\linewidth]{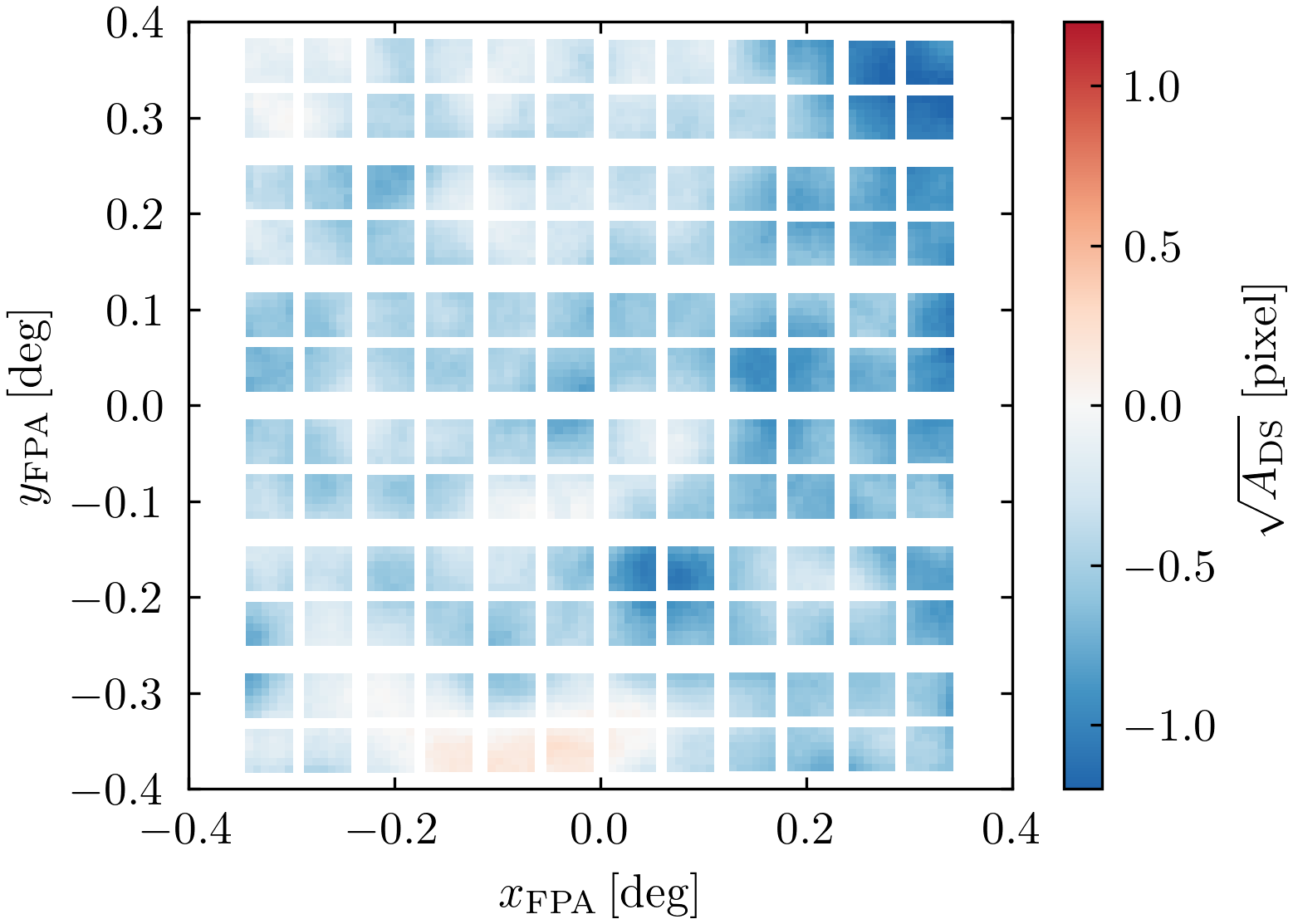}
    \caption{
    Amplitude of the final template \sqrtADS{} as a function of position in the focal plane array (FPA).
    We bin \sqrtADS{} in a grid of 6$\times$6 pixels per CCD-quadrant and take the biweight of stars as summary statistic for the respective pixel.
    The gaps between the four quadrants appear substantially larger than in the actual footprint because we include only stars located more than 100 pixels from the detector edges, in order to avoid truncation of the diffraction spikes.
    The peak-to-valley range of \sqrtADS{} is $1.51\,$pixels, corresponding to a defocus $\Delta z_\mathrm{star}$ range of $4.51\,\micron$.
    The sign of $\sqrt{A_\mathrm{DS}}$ is determined by the orientation of the triangle enclosed by the three diffraction spikes.
    }
    \label{fig:fov_template_fiducial}
\end{figure}

\begin{figure*}
    \centering
    \includegraphics[width=1.0\linewidth]{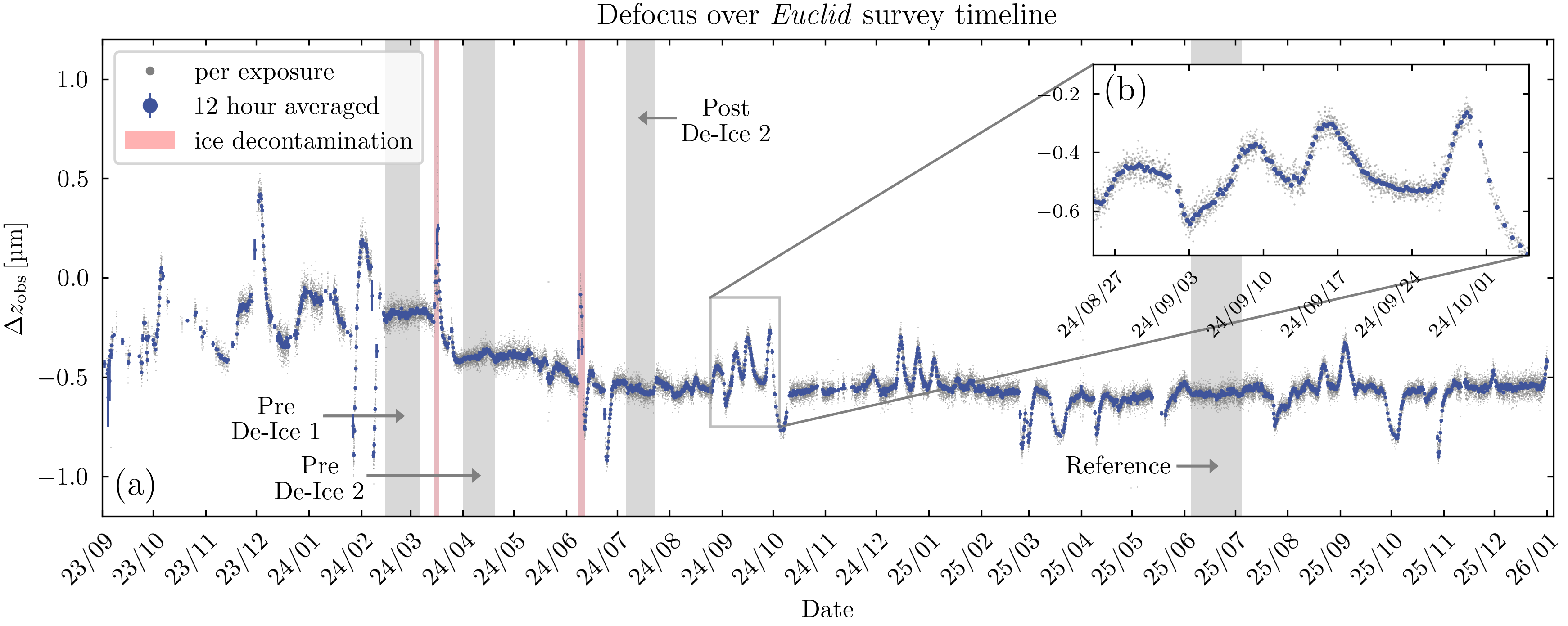}
    \caption{
    Evolution of the telescope defocus over the full survey timeline (a) and a zoom-in into the month of November 2024 (b). 
    Grey points show the inferred defocus per individual exposure and blue points show independent 12 hour averages, including the standard error of the mean. Grey shaded bands mark specific time periods used throughout this work, and red shaded bands indicate ice decontamination campaigns.
    \label{fig:DefocusFullSurvey}}
\end{figure*}

The estimated uncertainty of 0.022\,\micron{} is only achievable if \sqrtADS{} is drawn from the same distribution across the full FoV. 
This assumption is not true because the general FoV-dependent amplitude of \sqrtADS{} can show non-trivial structure due to the underlying WFE of the telescope optics.
If unaccounted for, this position-dependency leads to noisier defocus estimates due to sampling effects on the CCD.
However, to first order, the change in \zstar{} is constant across the field of view (\Cref{apdx:non_uniformity_of_defocus_change}).
Thus, by subtracting the baseline defocus distribution, we expect individual stars to be approximately independent and identically distributed random variables of \sqrtADS{}.
This assumption breaks down once the underlying background template changes, for example, due to a significant change in the telescope state.
We will examine this in more detail in \cref{sc:Results-deicing-influence}. 

We adopt an iterative process to infer accurate and precise defocus for each \Euclid-VIS instrument long exposure\footnote{The \Euclid observation sequence is described in detail in \citet[][figure 8 in particular]{Scaramella-EP1}.}:
\begin{enumerate}
    \item We first run the spike detection algorithm for stars with $12.22<m_G<15.86$, yielding \sqrtADS{} and auxiliary information like position in the image, \Gaia{} magnitude, and \Gaia{} photometric colour, obtained by crossmatching with the \Gaia{} DR3 release \citep{vallenari_gaia_2023}.
    \item As a noisy defocus estimate, we take the median of \sqrtADS{} per exposure and convert it to a defocus using \cref{eq:diffspike_to_defocus}. 
    \item We select a period of exposures that show little variation in defocus, in this case 2025/06/05--2025/07/05, henceforth denoted as `reference', also highlighted in \cref{fig:DefocusFullSurvey}.
    The \num{274589} bright stars from all \num{2207} exposures in this period are combined into a single high resolution footprint. 
    We then bin the \sqrtADS{} values of the stars in 6$\times$6 grid per CCD quadrant and use the biweight\footnote{Using the median here would cause a significant amount of stars in the population to be exactly corrected to zero, which distorts later population analyses.} \citep{kafadar_efficiency_1983} as a baseline value per cell to be corrected for later on. This template is shown in \cref{fig:fov_template_fiducial}.
    \item We subtract the baseline \sqrtADS{} obtained in step 3 from all exposures, take the median of \sqrtADS{} and convert it to defocus using 
    \begin{align}
        \label{eq:diffspike_correction}
        \Delta z = a\,\sqrt{A_\mathrm{DS}} + b_\mathrm{ref}\,,
    \end{align}
    where $a$ is the slope from \cref{eq:diffspike_to_defocus} and $b_\mathrm{ref}$ is the median inferred defocus in the stable time period from step 2. As a result, we obtain a summary defocus value for \num{52268} exposures between 2023/08/03 and 2026/01/12 under the assumption of a constant offset across the FoV.
\end{enumerate}
In principle, stopping after step 2 is already sufficient to monitor the telescope to relatively high precision. 
The relative change of defocus is encoded in \sqrtADS{} and one could already get a comprehensive overview of the telescope behaviour without its conversion to $\Delta z$.
However, in this work we want to emphasise the power of this method and maximise the information we can extract with it, especially with a telescope like \Euclid that is remarkably well suited for the described procedure.

\section{\label{sc:Results}Results}

\subsection{\label{sc:Results-Temporal Evolution of Defocus}Temporal evolution of defocus}

\begin{figure*}[t]
    \centering
    \includegraphics[width=1\linewidth]{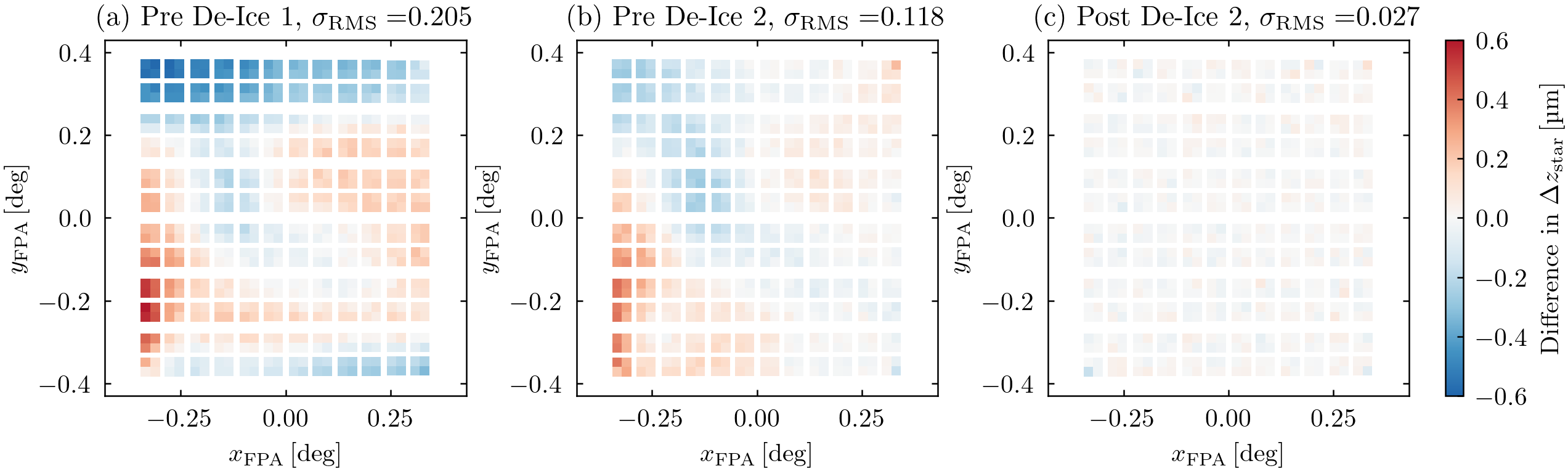}
    \caption{
    FoV-dependent defocus difference between exposures binned in the reference time frame and subsequent de-icing campaign time frames.
    (a) compares the reference and pre first decontamination, (b) reference and pre second decontamination, and (c) reference and post second decontamination. The individual templates are first binned in a 2$\times$2 grid per CCD quadrant. After subtracting the reference template, the images are additionally median subtracted to highlight non-constant offsets. The different time frames in which the exposures are binned are specified in \cref{tab:decontamination_defocus_evolution}.
    }
    \label{fig:fov_deicing_influence}
\end{figure*}

The inferred field-averaged defocus over time is displayed in \cref{fig:DefocusFullSurvey}. 
The general defocus consistently varies around a value of about $-0.56\,\micron$ since July 2024. 
However, deviations of the order of 0.3\,\micron{} are apparent within a time span of a week. 
These variations are highly correlated with thermal variation of optical elements on board of the \Euclid telescope (Euclid Collaboration: Whittam et al. 2026, in prep.) and are two orders of magnitude smaller than the thermal defocus shifts of the \HST.
% \citep{DR1-TP021} -> exchange above 

Short one-day gaps are typically caused by excess cosmic rays from solar flares and X-ray events, rendering this method unusable.
Another gap corresponds to observations of the Small Magellanic Cloud, causing the spike finder to fail due to an overabundance of bright sources close to individual stars.
% Apart from the above, some short one-day gaps appear, for example the end of 2024/10 and beginning of 2025/04. These correspond to exposures during solar flares that are highly polluted by cosmic rays and X-ray events and are thus unusable for this method.
% Further gaps in 2024/11 are caused by observations of the Small Magellanic Cloud, causing the spike finder to fail due to an overabundance of bright sources close to individual stars.
However, these constitute a negligible fraction of the vast number of exposures displayed in \cref{fig:DefocusFullSurvey}.

Disregarding the aforementioned outlier effects and short term thermal variations, the defocus of the \Euclid VIS instrument remains remarkably stable since July 2024.
Prior to that, the \Euclid telescope underwent two ice decontamination campaigns which we discuss in more detail in \cref{sc:Results-deicing-influence}.
We further quantify the spread of defocus using the 16th and 84th percentile in addition to the median within the reference timeframe from 2025/06/05 to 2025/07/05, amounting to \linebreak $\Delta z_\mathrm{obs,stable} = (-0.585{\pm0.027})\,\micron$.
This agrees well with the derived expected statistical uncertainty of 0.022\,\micron{} from \cref{sc:Methodology-Expected Uncertainty}, considering the slight but apparent temporal evolution of the reference period visible in \cref{fig:DefocusFullSurvey}.
If the intrinsic FoV dependency of \zstar{} is not accounted for, then within the same reference timeframe \linebreak $\Delta z_\mathrm{obs,stable}^{\rm uncorrected} = (-0.585{\pm0.087})\,\micron$.
Thus, even though the \zstar{} baseline subtraction does not change the average \zobs{}, the uncertainty increases by a factor of 3.22.

We conclude that, disregarding outliers, the inferred defocus of the \Euclid VIS instrument is not limited by systematic errors like residual FoV dependency, but rather by the statistical lower bound of the spike detection method itself.
Furthermore, the largely stable defocus validates a design that enables accurate modelling of the PSF over most of the survey timeline.
Yet, looking at the field-averaged defocus change per exposure only tells part of the story. 
Figure \labelcref{fig:DefocusFullSurvey} is based on the assumption that the defocus changes uniformly across the focal plane after subtracting the baseline FoV dependence given in \cref{fig:fov_template_fiducial}.
In \Cref{apdx:non_uniformity_of_defocus_change}, we verify that this is a reasonable assumption for the major part of the survey.
However, in the following, we highlight instances where this breaks down and discuss consequences for the PSF modelling.

\subsection{\label{sc:Results-deicing-influence}Influence of the de-icing campaign on the general wavefront error}

A non-uniform change of defocus might be an indicator of significant change in the general WFE behaviour of the spacecraft. As a case in point, we demonstrate the noticeable impact of the de-icing campaign on the general FoV template of \zstar{}. 
Specific mirrors were heated up to evaporate ice deposits on two separate occasions (Euclid Collaboration: Schirmer et al. 2026, in prep.).
% \citep{DR1-TP026} -> exchange above
This ice built-up poses a common issue for space telescopes \citep{haemmerle_cassini_2006, brieda_numerical_2022,Schirmer-EP29}.
The first decontamination was conducted between the 14th and 17th of March 2024, where FoM3 and M3\footnote{The exact labelling of the optical path design is described by Fig. 6 in \citet{EuclidSkyOverview}.} were respectively heated up by roughly 30$\,$K.
After a reoccurrence of ice, a second decontamination was executed on FoM3 alone beginning June 8th 2024. 
Both of these events are highlighted in red in \cref{fig:DefocusFullSurvey}.
In this section, we examine the impact on the baseline defocus FoV behaviour.

\begin{table}[ht]
\renewcommand{\arraystretch}{1.5}
\caption{Summary of the time frames used for reviewing change in telescope state after ice decontaminations.}
\label{tab:decontamination_defocus_evolution}
\resizebox{0.5\textwidth}{!}{%
\begin{tabular}{lccc}
\toprule
Period & Timeframe & $\widetilde{\Delta z}_\mathrm{obs}\,[\micron]$ & $N_\mathrm{obs}$ \\
\midrule
Pre De-Ice 1 & 2024/02/15--2024/03/07 & $-0.179^{+0.035}_{-0.036}$ & 1566 \\
Pre De-Ice 2 & 2024/04/01--2024/04/20 & $-0.394^{+0.029}_{-0.025}$ & 1380 \\
Post De-Ice 2 & 2024/07/06--2024/07/23 & $-0.567^{+0.031}_{-0.029}$ & 1196 \\
Reference & 2025/06/05--2025/07/05 & $-0.577\pm0.027$ & 2207 \\
\bottomrule
\end{tabular}
}
\tablefoot{The column $\widetilde{\Delta z}_\mathrm{obs}$ depicts the median defocus with the 84th and 16th percentile as the respective upper and lower bound. $N_\mathrm{obs}$ is the total number of used exposures per time frame.}
\end{table}
We combine exposures of temporally stable regions in defocus prior to the first and second decontamination (Pre De-Ice 1 and 2, respectively), post second decontamination \linebreak(Post De-Ice 2) and a reference time frame to visualise the change.
The chosen time frames are specified in \cref{tab:decontamination_defocus_evolution} and highlighted in \cref{fig:DefocusFullSurvey}.
The reference time frame is the same temporally stable series of exposures used to create the baseline defocus FoV template shown in \cref{fig:fov_template_fiducial}.
It begins a year after the second ice decontamination campaign and is thus fully independent of the other time series considered in this section.

Analogously to \cref{sc:Methodology-Iterative defocus calculation}, we create the corresponding FoV templates in each time frame by binning all stars in a 2$\times$2 grid per CCD quadrant.
We further convert \sqrtADS{} to $\Delta z$ using the known relation and display the FoV-dependent difference between reference and the other selected observations in \cref{fig:fov_deicing_influence}. Panels (a) and (b) show a noticeable difference in the FoV-dependent defocus component with respect to the reference. The root-mean-square (RMS) variation across the focal plane in the Pre De-Ice 1 period is $\sigma_\mathrm{RMS}(\Delta z_\mathrm{star})=0.205\,\micron$ (corresponding to $\sigma_{\mathrm{RMS}}(\sqrt{A_\mathrm{DS}})=\,0.069\,\mathrm{pixel}$) and shows significant structure.
For comparison, the RMS variation in the Post De-Ice 2 observation is only $\sigma_\mathrm{RMS}(\Delta z_\mathrm{star})=0.027\,\micron$.

This change cannot be explained by slight defocus non-uniformities that we found in \Cref{apdx:non_uniformity_of_defocus_change}.
The second ice decontamination was motivated by a rapid throughput degradation beginning a week after the first decontamination, suggesting substantial ice build-up during that time (Euclid Collaboration: Schirmer et al. 2026, in prep.).
% \citep{DR1-TP026} -> exchange above
However, the Pre-De-Ice 2 time series does not contain exposures within a month of the second de-icing campaign; those most heavily affected by ice.
A visual inspection of single day concatenated defocus FoV distributions within Pre-De-Ice 1 and Pre-De-Ice 2 shows no obvious temporal evolution, making ice build-up itself for the apparent difference in inferred defocus unlikely.
We conclude that the general WFE had to undergo significant, irreversible changes during each decontamination campaign to have such an impact on the \zstar{} baseline.
This is further accentuated by the fact that the relation between \zobs{} and temperature itself likely changed during that time (Euclid Collaboration: Whittam et al. 2026, in prep.), indicating a lasting alteration of the optical path. 
% \citep{DR1-TP021} -> exchange above
Where this change stems from remains unclear, but given the relatively large increase in temperature during contamination, the origin is possibly mechanical.
Only the heating element deposited on FoM3 was used during the second ice decontamination, while M2 and close-by elements remained cold.
Therefore, heating any optical element can non-trivially affect the focal length.
An alteration of similar severity has not occurred since the second decontamination campaign.

\subsection{\label{sc:Results-Temporal Evolution of FoV Template Uniformity}Temporal evolution of defocus FoV baseline distribution}

\begin{figure}
    \centering
    \includegraphics[width=1\linewidth]{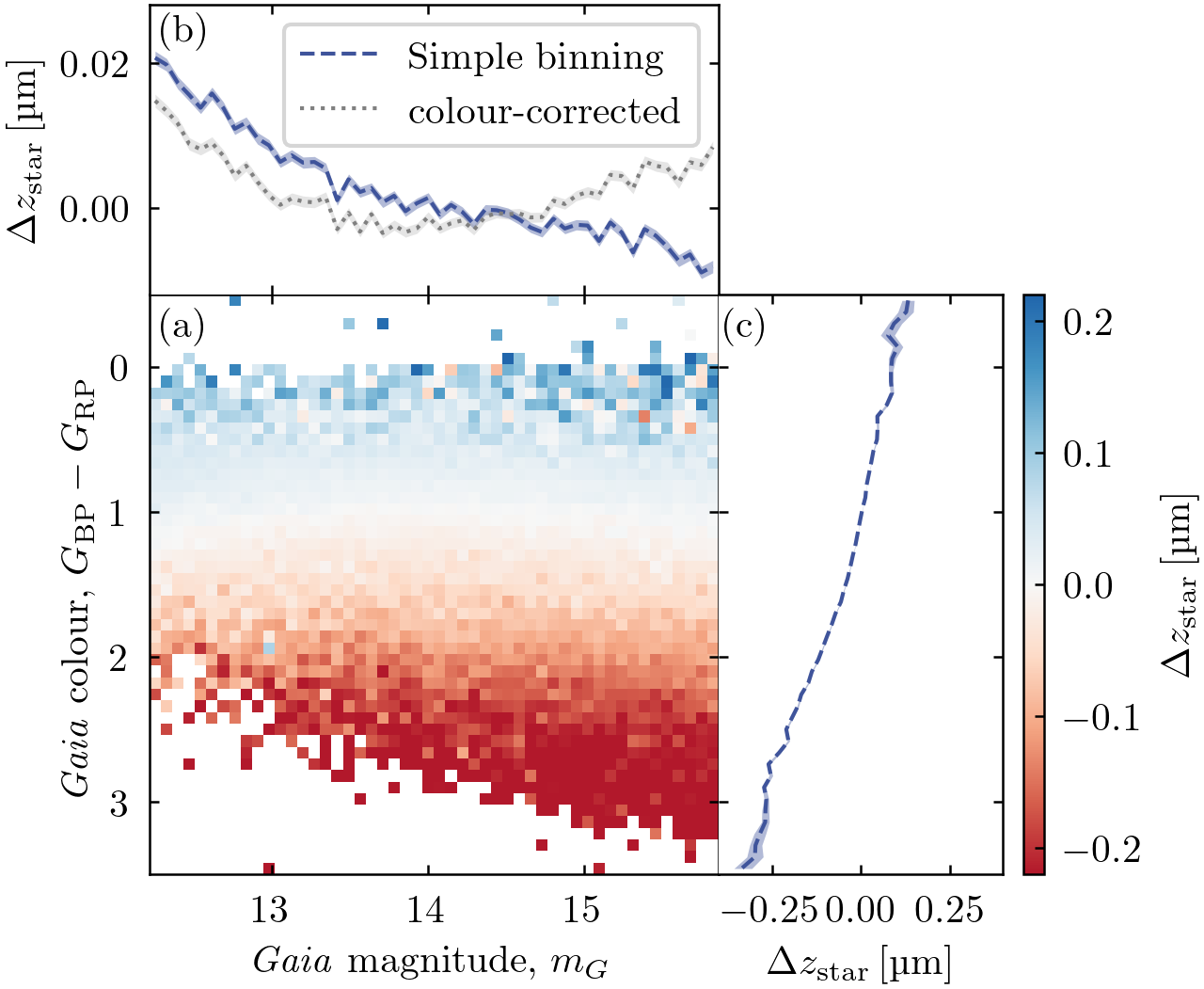}
    \caption{
    Defocus dependence on \Gaia{} $G$-band and colour.
    (a) shows the two dimensional histogram of the defocus dependency on \emph{Gaia} $G$-band magnitude and \emph{Gaia} colour defined as the difference between blue and red photometric band, with bluer stars towards the top. The colour-coded values are obtained by taking the median defocus value of all stars within the colour-magnitude bin.
    (b) and (c) show the defocus marginalised over \emph{Gaia} magnitude and \emph{Gaia} colour as blue dashed lines, respectively. The values are obtained by taking the median per bin and the shaded regions by dividing the 16th and 84th percentile by the number of stars in each bin, analogous to a standard error of the mean.
    Additionally, in (b) we show the \zstar{}--$m_G$ relation when correcting all defoci towards $G_{\rm BP}-G_{\rm RP}=1.02$ using the relation from (c) as grey dotted lines, effectively removing colour-dependence.
    Note that \zstar{} is an order of magnitude more sensitive to colour variations than to magnitude variations, both of which are smaller than the statistical uncertainty of the detection method [$\sigma(z_\mathrm{star})\approx0.247\,\micron$] for typical \Euclid exposures.
    }
    \label{fig:defocus_star_properties}
\end{figure}

% \begin{figure}[ht]
%     \centering
%     \includegraphics[width=1\linewidth]{figures/polynomial_coefficients_time_series_order1.png}
%     \caption{
%     Temporal evolution of the first order polynomial coefficients fitted to the residual FoV $\Delta z$ after subtracting the fiducial template shown in \cref{fig:fov_template_fiducial}. The $c_i$ are defined according to \cref{eq:polynomial_second_order_fit_defocus}.
%     }
%     \label{fig:temporal_evolution_polynomial}
% \end{figure}

In \cref{sc:Results-deicing-influence}, we established that the diffraction spike method is precise enough to trace changes in the underlying WFE.
In this section, we investigate the long-term temporal evolution in WFE, as represented by the \zstar{} distribution across the FoV. 
We quantify the evolution with Chebyshev polynomial coefficients that are fitted to the difference between combined exposures within stable periods and the baseline \zstar{} FoV distribution obtained from the reference time frame.
Assuming a linear evolution of polynomial coefficients, we trace the drift in the baseline \zstar{} FoV distribution over the full survey timeline since the second ice decontamination campaign.

We select stable seven day defocus time frames by first averaging \zobs{} in 12 hour intervals and then removing the weeks in which the averaged \zobs{} varies by more than 0.1\,\micron. 
Analogously to \cref{sc:Results-deicing-influence}, we collect all bright stars per stable seven day window into a single observation, bin \zstar{} into a 2$\times$2 grid per CCD-quadrant, subtract the reference FoV baseline and fit a 2d polynomial in Chebyshev base defined by
\begin{equation}
    \label{eq:2d_polynomial_chebyshev}
    \Delta z_\mathrm{star}(\hat{x},\hat{y}) = \sum_{i=0}\sum_{j=0}\,c_{ij}\, T_i(\hat{x})\,T_j(\hat{y})
\end{equation}
to the residual defocus, where the $T_{i/j}$ are Chebyshev polynomials of the first kind, and $\hat{x}\coloneqq x_\mathrm{FPA}/(0.344\,\rm deg)$ and $\hat{y}\coloneqq y_\mathrm{FPA}/(0.383\,\rm deg)$, resulting in $\hat{x},\,\hat{y}\in [-1,\,1]$ across the FoV.
In practice, we only consider terms up to second order in $\hat{x},\,\hat{y}$, effectively suppressing small-scale fluctuations that are more prone to noise due to the binning scheme.
We are not interested in accurately describing the FoV residuals, but rather in utilising them to detect any significant deviation from a constant offset for which a low-order polynomial fit already suffices.
As a result, we obtain the residual defocus FoV behaviour encoded in $c_{ij}$ as a function of time.
Non-zero coefficients indicate then a change with respect to the reference taken between 2025/06/05 and 2025/07/05.
% The evolution of the first order polynomial coefficients is depicted in \cref{fig:temporal_evolution_polynomial}, where we display the temporal evolution of the first order polynomial coefficients labelled according to \cref{eq:polynomial_second_order_fit_defocus}. 
% Large outliers are caused by a lack of data points per time frame or the same anomalies as for the defocus variation discussed in \cref{sc:Results-Temporal Evolution of Defocus}.
% Despite these irregularities, there is no apparent systematic evolution of the polynomial coefficients over time.
% Furthermore, $c_1$ and $c_2$ fluctuate consistently around zero. 
% For clarity, we only display the first order coefficients in \cref{fig:temporal_evolution_polynomial}.
To quantify the temporal evolution of the coefficients, we perform a linear fit of the form $c_{ij}(t)=k\,t+c_{ij,0}$ for the individual $c_{ij}$ that would reveal slow drifts in time $t$. 
The results are shown in \cref{tab:polynomial_fit_temporal_evolution}.

\begin{table}[ht]
\centering 
\caption{
Temporal evolution of the second-order polynomial coefficients describing the FoV difference between exposures collected in subsequent stable seven day periods and the baseline FoV behaviour, along with the fit uncertainty.
}
\label{tab:polynomial_fit_temporal_evolution}
\resizebox{0.4\textwidth}{!}{%
\begin{tabular}{l
                S[separate-uncertainty, table-format=1.3(2)]
                S[separate-uncertainty, table-format=1.2(2)]}
\hline \hline
\noalign{\vskip 1pt}
Coefficient & {$k\,[10^{-4}\,\micron\,\mathrm{day}^{-1}]$} & {$c_{ij,0}\,[10^{-3}\,\micron]$} \\
\hline
\noalign{\vskip 1pt}
$c_{00}$ & 0.030 \pm 0.018 & 18.71 \pm 0.64 \\
$c_{10}$ & -0.003 \pm 0.025 & 2.09 \pm 0.88 \\
$c_{01}$ & -0.017 \pm 0.025 & -0.33 \pm 0.88 \\
$c_{20}$ & 0.031 \pm 0.024 & -1.90 \pm 0.86 \\
$c_{11}$ & -0.014 \pm 0.043 & 2.4 \pm 1.5 \\
$c_{02}$ & -0.031 \pm 0.024 & 1.70 \pm 0.84 \\
\hline
\end{tabular}
}
\tablefoot{The slope and intersect are defined by $c_{ij}(t)=k\,t+c_{ij,0}$, where the coefficients are labelled according to \cref{eq:2d_polynomial_chebyshev}.}
\end{table}

We set the zero-point in time to 2024/07/01 00:00:00.
The coefficients exhibit a small evolution over time, that agrees with zero within $2\sigma$. Extrapolating the trend from \cref{tab:polynomial_fit_temporal_evolution} over six years causes an RMS error increase of \linebreak $\sigma_\mathrm{RMS}(\Delta z_\mathrm{star})=0.0091_{-0.0044}^{+0.0076}\,\micron$ per exposure.
A change like that would be barely detectable and is much smaller than the temporal fluctuations seen in \cref{fig:DefocusFullSurvey}.
At the current rate, any long-term drift in the defocus WFE post second ice decontamination is negligible.

\subsection{\label{sc:Results-star-defocus-properties}Dependency of inferred star defocus on star colour and magnitude}

By themselves, the diffraction spike shifts introduced by defocus are achromatic (see \Cref{apdx:A}).
Therefore, a dependence of the inferred defocus on the star colour indicates mixing with other WFE that are chromatic.
In this section, we examine the relation between our inferred defocus estimate and star properties like colour and magnitude to expose any such couplings. 

We do so by collecting all stars across all exposures between 2024/07/01 and 2026/01/12 and correcting their defocus according to the reference FoV template from \cref{fig:fov_template_fiducial}, as well as subtracting the biweight defocus per exposure. 
In doing so, we should obtain a sample of stars with $\langle\Delta z_\mathrm{star}\rangle_\mathrm{biweight}=0$. 
We apply a colour filter, removing stars outside $-0.5<G_\mathrm{BP}-G_\mathrm{RP}<3.5$, corresponding to $0.01\%$ of the sample. 
The lower bound is motivated by the validity of a flux correction for bright stars, whereas the upper bound avoids stars with large $G_\mathrm{BP}-G_\mathrm{RP}$ flux excess, often caused by blending \citep{riello_2021_gaia}.
The final catalogue contains \num{5308816} stars across \num{37524} exposures.

We bin the stars in 50 equally-spaced colour and magnitude bins and represent the star defocus per bin by taking the median as summary and the 16th and 84th percentile divided by the square root of number of stars as uncertainty.
The results are displayed in \cref{fig:defocus_star_properties}.
\zstar{} is an order of magnitude more sensitive to star colour than to magnitude, as is apparent in \cref{fig:defocus_star_properties}, panel (c).
90\% of the stars lie between $G_\mathrm{BP}-G_\mathrm{RP}\in [0.70,\,1.73]$, covering a defocus range of $\Delta z_\mathrm{star}\in[0.029,\,-0.071]\,\micron$.
The defocus-colour dependence is clearly detectable and indicates coupling to chromatic sources of WFE. 
A likely candidate for such a source is the dichroic coating used to split the light beam entering the \Euclid telescope for the VIS and NISP detectors \citep{EuclidSkyOverview}.
It consists of dielectric layers that induce a highly chromatic reflectivity response \citep{baron_measurement_2022} that needs to be taken into account for accurate PSF modelling (Euclid Collaboration: Miller et al. 2026, in prep.).
% \citep{DR1-TP018} -> exchange above
A dichroic mirror uses interference by means of reflection in thin layers to create a wavelength-dependent reflectivity that naturally induces wavelength-dependent phase-shifts as well, such that $W_{\rm E}=W_{\rm E,defocus}(r^2)+W_{\rm E,dichroic}(x,y,\lambda)$. The \linebreak exact wavelength response is highly setup dependent, but produces high wavefront errors in particular for $\lambda>800\,\si{\nano\metre}$ for the \Euclid optical system \citep{venancio_2016_coating}, making the shift in diffraction spikes incoherent.
However, we stress that, even though we clearly detect a colour-dependence of \zstar{}, the chromatic variations are smaller than the uncertainty of $\sigma(\Delta z_\mathrm{star})=0.247\,\micron$ for typical stellar colour fluctuations of bright stars \citep{E-Anselmi}.
Nevertheless, small biases are expected when pointing at fields that, for example, contain peculiar stellar populations.

The dependence of \zstar{} on the star magnitude in \cref{fig:defocus_star_properties}, panel (b), is an order of magnitude weaker.
We display two relations, one obtained from simple binning in \Gaia{} $G$ band magnitude, and one from correcting all \zstar{} towards $G_\mathrm{BP}-G_\mathrm{RP}=1.02$ (mean colour of the sample) prior to binning using the relation obtained from panel (c).
This colour-corrected defocus-magnitude relation is less monotonic than the uncorrected counterpart, albeit of similar magnitude.
The reason behind this magnitude-dependence is at least in part caused by the detection method itself:
As discussed in \cref{sc:Methodology-Expected Uncertainty}, oversaturated stars can show excessive bleeding, causing bright vertical streaks in the CCD readout direction.
One of the diffraction spikes is only rotated from the vertical by approximately 10$^\circ$ and thus close to bleeding streaks.
This can interfere with the spike detection for stars on the bright end.
On the faint end, the spike brightness moment dipole could systematically shift if the brightness profile is not symmetric per spike, causing non-trivial influence of the relative increased noise-floor for faint spikes.
We also expect some of the magnitude dependence to be driven by the aforementioned colour dependence due to the correlation between colour and brightness \citep{babusiaux_gaia_2018}, alone from the fact that red stars are less abundant for lower magnitudes as apparent in panel (a).
\\[1em]
\indent In summary, we detect chromaticity within our defocus estimate, likely driven by the dichroic semi-permeable mirror coating that exhibits a wavelength-dependent response.
However, \citet{E-Anselmi} show that 12 hour averaged mean colour varies within $\delta(\langle G_\mathrm{BP}-G_\mathrm{RP}\rangle)<0.1$, corresponding to a defocus change of $\mathcal{O}(0.01\,\micron)$ which is subdominant in relation to the detection method uncertainty and is unlikely to affect the field averaged defocus \zobs{} significantly.

\subsection{\label{sc:Results-Correlation of Defocus with PSF core size}PSF stability in relation to the \Euclid PSF size requirements}

The PSF size is strongly correlated with defocus, and thus, our defocus estimate presented here. 
Finding this (FoV-dependent) relation enables us to compare the estimated defocus variations to the \Euclid PSF size requirement of $\left|\Delta R_\mathrm{PSF}^2/R_\mathrm{PSF}^2\right|<10^{-3}$ \citep{cropper_defining_2013}. 
Our goal in this section is twofold. First, we want to get a rough estimate on how strong a mis-modelled defocus perturbs the PSF size, essentially asking what defocus threshold $\Delta z_{\rm obs,thr}$ can be tolerated without needing to adjust the model. Second, we want to examine how our per exposure defocus precision compares to this allowed defocus variation.

The results depend on the choice of the fiducial (in-focus) PSF size, $R^2_{\rm PSF,true}$, which is highly FoV-dependent. 
As mentioned in \cref{sc:Background-Relation between defocus and PSF size}, we expect a quadratic relation between $R_\mathrm{PSF}^2$ and \zstar{}, and by extension, between $\Delta R_\mathrm{PSF}^2/R_\mathrm{PSF}^2$ and \zstar{} that differs from CCD-quadrant to quadrant. 
Any misestimation of $R^2_{\rm PSF,true}$ propagates quadratically into the corresponding defocus threshold.
From the data alone, we have no access to $R^2_{\rm PSF,true}$.
We limit ourselves in this section therefore to two estimates; we derive the allowed defocus threshold based on an idealised scenario where $R^2_{\rm PSF,true}$ is at the minimum of the quadratic curve, and one based on a more realistic scenario where $R^2_{\rm PSF,true}$ is chosen as the most common $R^2_{\rm PSF}$ per quadrant.
Note that we do not expect these two to coincide.
The former sets the lower-bound of the defocus-size response, while the latter threshold corresponds to perturbations around the nominal defocus of the spacecraft.
Naturally, other PSF metrics like ellipticity are pivotal for weak lensing analyses as well, but will not be explored in this work.
More discussion about them can be found in Euclid Collaboration: Whittam et al. (2026, in prep.).
% \citet{DR1-TP021} -> exchange above

We measure $R_\mathrm{PSF}^2$ of all stars with \Gaia{} magnitude $19<m_G<21$ from raw \Euclid VIS exposures observed since July 2024 using \cref{eq:quadrupole_brightness_moments} with a Gaussian weight function ($\sigma=\ang{;;0.25}$), ignoring the brighter-fatter-effect due to its subdominance in the chosen magnitude range \citep{Q1-TP002}. 
The stars used for the defocus inference with $12.22<m_G<15.86$ constitute a significantly different stellar population. 
To reduce chromatic differences, we separately bin $R_\mathrm{PSF}^2$ and \zstar{} into 50 equal-width $G_\mathrm{BP}-G_\mathrm{RP}$ bins between $-0.5<G_\mathrm{BP}-G_\mathrm{RP}<3.5$ to obtain a colour-size relation. 
We then remove the mean chromaticity by subtracting the colour-dependent offset relative to $(G_\mathrm{BP}-G_\mathrm{RP})_\mathrm{ref}=1.02$. 
We hence assume that a given $\delta(G_\mathrm{BP}-G_\mathrm{RP})$ approximately induces a fixed $\delta R_\mathrm{PSF}^2$ or $\delta(\Delta z_\mathrm{star})$.
Without this correction, the mean reduced $\chi^2$ from the quadratic fits described below across all quadrants increases from 3.36 to 7.17, using $N_\mathrm{dof}=15$ degrees of freedom.\footnote{Note that a reduced $\chi^2$ of 3.36 still indicates a bad model/underestimated errors. Our binning implicitly assumes that \zstar{} and $R^2_{\rm PSF}$ are constant within each quadrant, which is not true.
Further subdividing the quadrants into a 2 $\times$ 2 grid reduces the average reduced $\chi^2$ to 1.67, while affecting the final allowed defocus range at the 5--10\% level.
Thus, the more impactful decision in this section is the choice of the `true' (in-focus) PSF size.}

Unlike in the previous sections, the total value of \zstar{} is relevant here, not just the relative change to the reference FoV distribution. 
However, we can still split the exposures in 20 equally spaced \zobs{} bins between $-0.85\,\micron$ and $-0.25\,\micron$ to obtain stellar populations with different defocus.
By equating the median $R^2_\mathrm{PSF}$ and \zstar{} per \zobs{} bin and quadrant, we obtain the desired size-defocus relation per CCD-quadrant as 
\begin{align}
    R_\mathrm{PSF}^2(\Delta z_\mathrm{star}) = a\,(\Delta z_\mathrm{star}-\Delta z_\mathrm{star,min})^2 +R_\mathrm{PSF,min}^2\,,
\end{align}
where $a$, $\Delta z_\mathrm{star,min}$ and $R_\mathrm{PSF,min}^2$ are free fitting parameters. 
Disregarding the first and last bin due to limited sample size, this amounts to $N_\mathrm{dof}=18-3$.

Therefore, we can estimate the maximum allowed defocus threshold $\Delta z_\mathrm{star,thr}$ as
\begin{align}
    \left|\Delta R_\mathrm{PSF}^2/R_\mathrm{PSF}^2\right|({\Delta z_\mathrm{star,thr}})=10^{-3}\,,
\end{align}
where we adapt the definition of the size bias (\Cref{eq: DeltaR2/R2}) to 
\begin{align}
    \label{eq:DeltaR2/R2 effective}
    \Delta R_\mathrm{PSF}^2/R_\mathrm{PSF}^2 \coloneqq \frac{R_\mathrm{PSF}^2(\Delta z_\mathrm{star})-R_\mathrm{PSF,true}^2}{R_\mathrm{PSF,true}^2}\,,
\end{align}
assigning $R_\mathrm{PSF}^2(\Delta z_\mathrm{star})$ as the modelled PSF size. The slope of the quadratic relation is smallest around $R_\mathrm{PSF,min}^2$, so we choose $R_\mathrm{PSF,true}^2=R_\mathrm{PSF,min}^2$ for a lower bound estimation on the size-defocus response.
Any other choice of $R_\mathrm{PSF,true}^2$ will increase the steepness of the response, leaving $\Delta z_\mathrm{thr}$ as a valid upper bound. 
$\Delta z_\mathrm{star,thr}$ shows little variation across the CCD-quadrants, allowing us to summarise it into 
\begin{align}
\langle\Delta z_\mathrm{star,thr}\rangle=\Delta z_\mathrm{obs,thr}=(0.1598\pm0.0025)\,\micron \,,
\end{align}
using the (standard error of the) mean. 
We stress that this is 7 times larger than the uncertainty per exposure of our method. 

Repeating this analysis by choosing the average $R_\mathrm{PSF}^2$ in the most populated \zobs{} bin per CCD-quadrant as fiducial $R_\mathrm{PSF,true}^2$ yields
\begin{align}
\langle \Delta z_\mathrm{star,thr} \rangle=\Delta z_\mathrm{obs,thr}=(0.0683\pm0.0024)\,\micron \,.
\end{align}
Effectively, this defocus threshold quantifies the allowed perturbation around the \emph{nominal} defocus that is acceptable within \Euclid requirements.

Such a change would still be detectable with a shift in diffraction spikes for a typical exposure that contains $\mathcal{O}(100)$ stars, and entire weeks of survey exposures show comparable spread in variability. 
In both cases, our defocus estimate can provide a stringent prior for the PSF model, potentially drastically cutting down on the numerical expense of fitting a physical PSF model. In particular, extended time periods during nominal survey operation would require no a-priori change in the model from a $R_\mathrm{PSF}^2$ requirement perspective alone (see \cref{fig:DefocusFullSurvey}). 
Of course, the effects of defocus on other PSF metrics, such as ellipticity and trefoil, are more tentative. 
Further discussion can be found in Euclid Collaboration: Whittam et al. (2026, in prep.).
% \citet{DR1-TP021} -> exchange above

In \cref{sc:Results-Temporal Evolution of FoV Template Uniformity} we examined the long-term temporal evolution of the FoV-dependent defocus \zstar{} and projected a change in constant offset of $\Delta c_{00}=(0.0066\pm0.0040)$\,\micron{} over six years. 
Such a long-term bias will be smaller than the defocus change allowed considering the PSF size requirements, even in the realistic case.
Therefore, in addition to dedicated PSF calibration observations, exposures from extended stable time periods from the nominal survey itself could be used to directly inform the PSF modelling effort.
This highlights the successful endeavour of creating an unprecedented stable cosmology telescope, allowing the Euclid Consortium to precisely control weak lensing measurement systematics with a PSF model fuelled by accurate and precise WFE estimates.

\section{\label{sc:Conclusion}Conclusions}

Obtaining unbiased cosmology from weak gravitational lensing requires a firm grasp on the \Euclid telescope point spread function. 
In this work, we present a fast method to investigate the evolution of its most variable wavefront error: the defocus.
We can trace image plane shifts in terms of secondary \emph{Euclid} mirror position to a precision of $0.022\,\micron$ per exposure by utilising the diffraction spikes of $\mathcal{O}(100)$ bright stars, corresponding to a peak-to-valley optical-path difference of \qty{0.75}{\nano\metre}.
This defocus is empirically obtained from \Euclid observations thanks to the non-mirror-symmetric M2 spider struts in the entrance pupil and does not require an a-priori PSF model.
Therefore, it can not only serve as a powerful monitoring tool for the \Euclid spacecraft itself, but also provides an independent metric for validating the \Euclid PSF model, which is crucial for cosmological inference with weak gravitational lensing.

Distinguishing between a total estimated defocus per exposure \zobs{} and an estimated defocus of individual stars across the focal plane \zstar{}, this enables us to explore a multitude of effects related to this wavefront error:
\begin{itemize}
    \setlength\itemsep{1em}
    \item In \cref{sc:Results-Temporal Evolution of Defocus} we showed the temporal evolution of the defocus over the entire survey timeline until 2026/01/12. After large variations during initial calibration and ice decontamination campaigns, the spacecraft has been remarkably stable, showing extended periods where the variation in defocus is three orders of magnitude smaller than breathing variations of the \HST. 
    The largest defocus variations after July 2024 are on the order of \zobs$=0.6\,\micron$ and highly correlated with the telescope temperature (Euclid Collaboration: Whittam et al. 2026, in prep.). 
    % \citep{DR1-TP021} -> exchange above
    We can therefore trace the influence of opto-mechanical fluctuations common for space telescopes, which are not only crucial for monitoring, but also instructive for PSF modelling \citep{rhodes_stability_2007, cropper_defining_2013, liaudat_point_2023}.
    \item We demonstrate in \cref{sc:Results-deicing-influence} that two unplanned ice decontamination campaigns in March 2024 and June 2024 caused significant, persisting changes to the underlying focal plane-dependent amplitude of defocus (\cref{fig:fov_deicing_influence}).
    In the 1.5 years after the second decontamination campaign, however, our inferred defocus changes to the 3.7\% level uniformly across the field of view during nominal survey exposures (\Cref{apdx:non_uniformity_of_defocus_change}).
    Since the \Euclid phase diversity calibrations happened prior to the decontamination in March 2024 \citep[section 4.3.2 therein]{EuclidSkyOverview}, the \Euclid VIS PSF model should be monitored closely regarding misestimation of the wavefront errors going forward.
    \item We further identify one-week intervals with a stable \zobs{} and fit a second-order two-dimensional polynomial to the residual \zstar{} with respect to a stable reference time frame (2025/06/05--2025/07/05), allowing us to trace changes in the WFE baseline over time.
    We found evidence of a long-term evolution of the defocus field of view-dependence at the 1.7$\sigma$ level  in \cref{sc:Results-Temporal Evolution of FoV Template Uniformity} since July 2024. However, the resulting defocus offset extrapolated over six years is two orders of magnitude smaller than changes caused by thermal variations, confirming the temporal stability of the \Euclid spacecraft.
    \item The shift in diffraction spikes due to defocus is a-priori achromatic. 
    However, we detect a slight correlation between \zstar{} and \emph{Gaia} photometric colour (\cref{sc:Results-star-defocus-properties}).
    We attribute this colour dependence to coupling of chromatic WFE induced by the dichroic mirror in the VIS optical path.
    The dependency of \zstar{} on \emph{Gaia} colour is comparable to the statistical uncertainty of the defocus detection method and could lead to detectable biases when observing fields with peculiar stellar populations.
    However, we stress that the 12 hour averaged stellar colour varies typically $\delta(\langle G_\mathrm{BP}-G_\mathrm{RP}\rangle)<0.1$ \citep{E-Anselmi}.
    This corresponds to a maximum defocus change of $\approx0.01\,\micron$ due to stellar field variations.
    The dependence of \zstar{} on \emph{Gaia G}-band magnitude is an order of magnitude smaller and is explicable by systematic effects arising from our diffraction spike finder method.
    \item In \cref{sc:Results-Correlation of Defocus with PSF core size}, we examined the PSF size variation due to a misestimated defocus and related this to the \Euclid requirement of $\left|\Delta R_\mathrm{PSF}^2/R_\mathrm{PSF}^2\right|<10^{-3}$. We find a defocus threshold of ${\Delta z}_\mathrm{obs,thr}=(0.1598\pm0.0025)\,\micron$ as an upper bound, and ${\Delta z}_\mathrm{obs,thr}=(0.0683\pm0.0024)\,\micron$ for perturbations around the nominal \Euclid defocus. 
    Extended weeks of nominal survey exposures exhibit smaller variation in both cases, again highlighting the stability of the \Euclid space telescope. The precision of the defocus estimate with diffraction spikes enables the use of tight priors when fitting the physical \Euclid PSF model.
    These stable periods themselves could potentially be used as PSF calibration fields, significantly increasing the amount of stable observations necessary for creating a precise PSF model.
    This is particularly important for the \Euclid shape-measurement pipeline, which relies on repeated accurate predictions of the PSF \citep{EP-Congedo}.   
    We stress, however, that the defocus threshold derived from size requirements does not necessarily apply to other PSF metrics, such as ellipticity.
    A more in-depth investigation of additional PSF metrics is presented in Euclid Collaboration: Whittam et al. (2026, in prep.).
    % \citet{DR1-TP021} -> exchange above
    \item The precision of $0.022\,\micron$ per exposure is driven by (a) the statistical uncertainty of the spike detection method, and (b), the number of usable stars per exposure. 
    A possible way to tackle both limitations at the same time is to choose magnitude-dependent parameters of the spike detection algorithm.
    This could take the form of a magnitude-dependent inner and outer mask radius or a more optimal weight function for the spike detection.
    However, we demonstrated that our current `one-size-fits-all approach' allows for detailed analysis of the defocus wavefront error evolution, and we leave these optimisations for future work.
    Furthermore, in this work, we use raw \Euclid VIS images that are polluted by cosmic rays. Proper masking and in-painting of cosmic rays will likely reduce the defocus outlier fraction, making the summary statistic more robust.

    \item We demonstrated the effectiveness of this defocus estimator for the \Euclid space telescope, which is fortuitously well equipped for the presented inference.
    The condition for this purely empirical method are thin, straight obscurations in the telescope entrance pupil that are asymmetric when mirrored perpendicular to individual vanes.
    With $\mathcal{O}$(100) bright stars per exposure, one can in principle infer the defocus of any telescope with this configuration, without using additional wavefront sensors. 
    Still, there are some caveats that need to be investigated.
    Naturally, an increasing number of struts will lead to more complex enclosed diffraction spike areas, making it more difficult (albeit not impossible) to assign an orientation to the shape, and by extension, the direction of defocus. 
    More in-depth discussion about the application to the \emph{Roman} Space Telescope, amongst others, can be found in \Cref{apdx:pupil_configurations}.
    Additionally, the assumption that \zstar{} changes uniformly across the FoV is a good approximation for the \Euclid spacecraft, but must be confirmed on a case-by-case basis.
    Should this be proven to be generally true, then our method provides a fast and precise estimator of the telescope defocus that can be used for monitoring and model fitting.
\end{itemize}

In conclusion, we obtain a precise and accurate estimator of the VIS instrument defocus wavefront error from apparent shifts in the diffraction spikes of bright stars.
We demonstrate the unprecedented stability of the \Euclid spacecraft, highlighting its capability to provide coherent galaxy shape measurements over its full survey timeline.
Continuous surveillance of the VIS PSF temporal evolution is a crucial step towards the goal of constraining cosmology using gravitational shear in the years to come.
The defocus inference with the presented method in this work provides a quick and accessible gateway to monitor the evolution of the \Euclid spacecraft.
However, we stress that this method is not limited to the Euclid Survey.
Our defocus estimator is PSF-model independent and, thus, easily generalisable to any (space) telescope with straight, thin, and non-mirror-symmetric obscuration in the optical path entrance pupil like the \emph{Roman} Space Telescope \citep{spergel_widefield_2015, akeson_wide_2019}.

\begin{acknowledgements}
%\AckERO  
We thank Sylvain Mottet and Divya Rana for their valuable feedback. DN and HHo acknowledge funding from the European Research Council (ERC) under the European Union’s Horizon 2020 research and innovation program (Grant agreement No. 101053992).
\AckEC  
\end{acknowledgements}

%
% Here comes the reference list, generated via bibtex from
% your bibfile my.bib and Euclid.bib. Please make sure that
% the same paper is not referenced twice, one from your my.bib
% file, and once from Euclid.bib.
%

\bibliography{Euclid, DR1, mylib} % add my.bib, containing your bibentry file 

%
% Now you can add appendices.
% In this example, the appendices are in one column mode.
% If that is not requires, comment out \onecolumn
% Note that appendices in A\&A come {\it after\/} the references.

\begin{appendix}
%\onecolumn %If you don't want single column for the Appendix, please
             %comment this out
  
\section{\label{apdx:A}Derivation of the relation between defocus and diffraction spike position}

\begin{figure}
  \begin{centering}
  \includegraphics[width=0.4\textwidth]{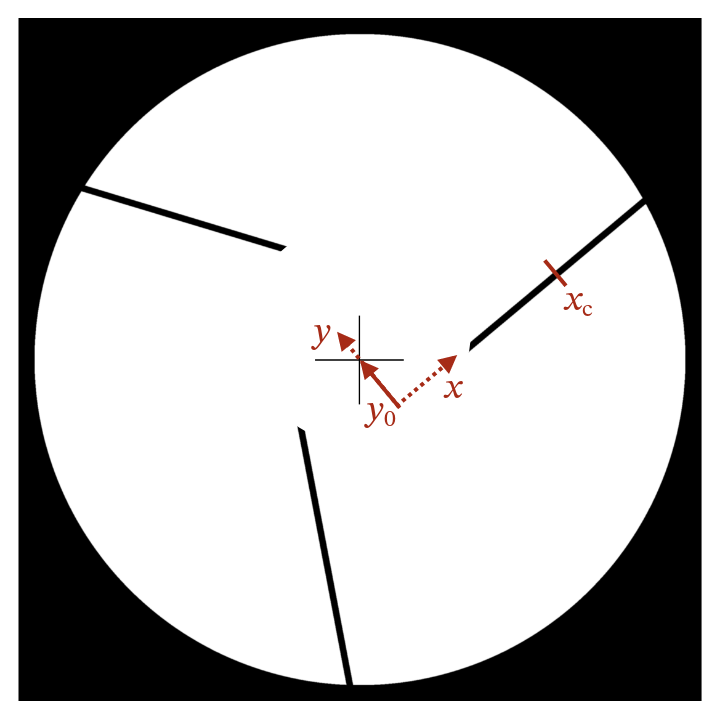}
  \caption{
    Schematic illustration of the location of the three M2 support struts in the \Euclid pupil plane, showing the coordinate system adopted for the right-most strut. The adopted $x, y$ coordinate system for each strut is defined with respect to its major and minor axes, respectively, $y_0$ is the perpendicular distance from the strut major axis to the phase centre (marked with a cross), and $x_{\rm c}$ is the centre of the strut in $x$-direction. The inner end of each strut terminates at the central M2 obscuration, which is omitted for clarity.
    \label{fig:schematic}}
  \end{centering}
\end{figure}

The \Euclid PSF has three pairs of diffraction spikes, arising from diffraction by the three spider struts supporting M2, with their location in the pupil plane shown in Fig.\,\ref{fig:schematic}.
Telescope defocus results in a wavefront phase error that varies quadratically with radius in the pupil plane. Locally at the strut obscuration in that plane, there is a radial phase gradient which, when Fourier transformed, results in a lateral shift of the diffraction spike.
Geometrically, one can ad hoc motivate this by interpreting each strut as a rectangular inverse aperture that creates a spike centred at its `centre of mass' extending orthogonal to the strut $x$-direction.
For an in-focus system, all parts of the pupil (including the spike centres) project to the same point on the detector. However, for a defocused system with the image plane a distance $\Delta z_\mathrm{eff}$ above the focus, the projected spike centres are offset from the in-focus diffraction pattern due to the conical beam by an amount $x_{\rm c}\,\Delta z_\mathrm{eff}/f$, for focal length $f$.
%This effect is the same as in the Bahtinov mask, used for focussing small telescopes.

The locations of the diffraction spikes may be used as a wavefront sensor of radial phase gradients in the pupil plane, at FoV locations wherever there is a bright star.
In a telescope with an even number of top-end spider struts that are symmetrically arranged, these lateral shifts are blurred together and thus not easily measurable, but owing to the M2 spider mirror asymmetry, the shifts are clearly detected in \Euclid images. According to Babinet's principle, the amplitude of the pupil plane complex electric field may be considered as a superposition of an open pupil, with no strut vignetting, minus a contribution from the struts alone \citep[page 424]{born_principles_1999}. This allows us to calculate the expected lateral shifts. 
The shadows of the struts are projected onto the entrance pupil. 
Given the dimensions in the \Euclid design, we may ignore near-field Fresnel
diffraction associated with that projection
and calculate the expected diffraction spike profiles,
assuming the struts are in the pupil plane, and hence assuming the Fraunhofer condition.

The diffraction pattern of each individual strut may then be calculated as the inverse Fourier transform of the strut, scaled by wavelength.
For this calculation, we may rotate and shift the $x-$ and $y-$coordinate axes in the pupil domain to align with the geometry of each strut.
The Fraunhofer angular diffraction pattern $f(u_x, u_y)$ as a function of direction cosines $u_x,u_y$ is then:
\begin{equation}
    \label{eq:fraunhofer_general}
    f(u_x,u_y) = \iint_{\rm pupil} {\rm e}^{{\rm i}k_xu_x}\,{\rm e}^{{\rm i}k_yu_y}
    \,{\rm e}^{{\rm i}\phi(k_x,k_y)}\,\diff k_x\,\diff k_y \, ,
\end{equation}
where $k_x = 2\pi x/\lambda$, $k_y=2\pi y/\lambda$ are the pupil plane wavenumbers along the
two strut axes and where $\phi(k_x, k_y)$ is the
wavefront phase as a function of pupil plane location \citep{born_principles_1999}.
Note that the detector position $x_i\in\{x,y\}$ is related to its direction cosine with the focal length $f$ via $u_{x_i}\approx x_i/f$ for small angles.

If we neglect the variation in phase across the width of the strut, assumed in the $k_y$ direction, the
$u_y$ dependence is the usual sinc function with length that depends on the strut width.  The profile across
the diffraction spike is then given by a 1D inverse Fourier transform on the $k_x$-axis.

\begin{figure}
  \begin{centering}
  \includegraphics[width=0.45\textwidth]{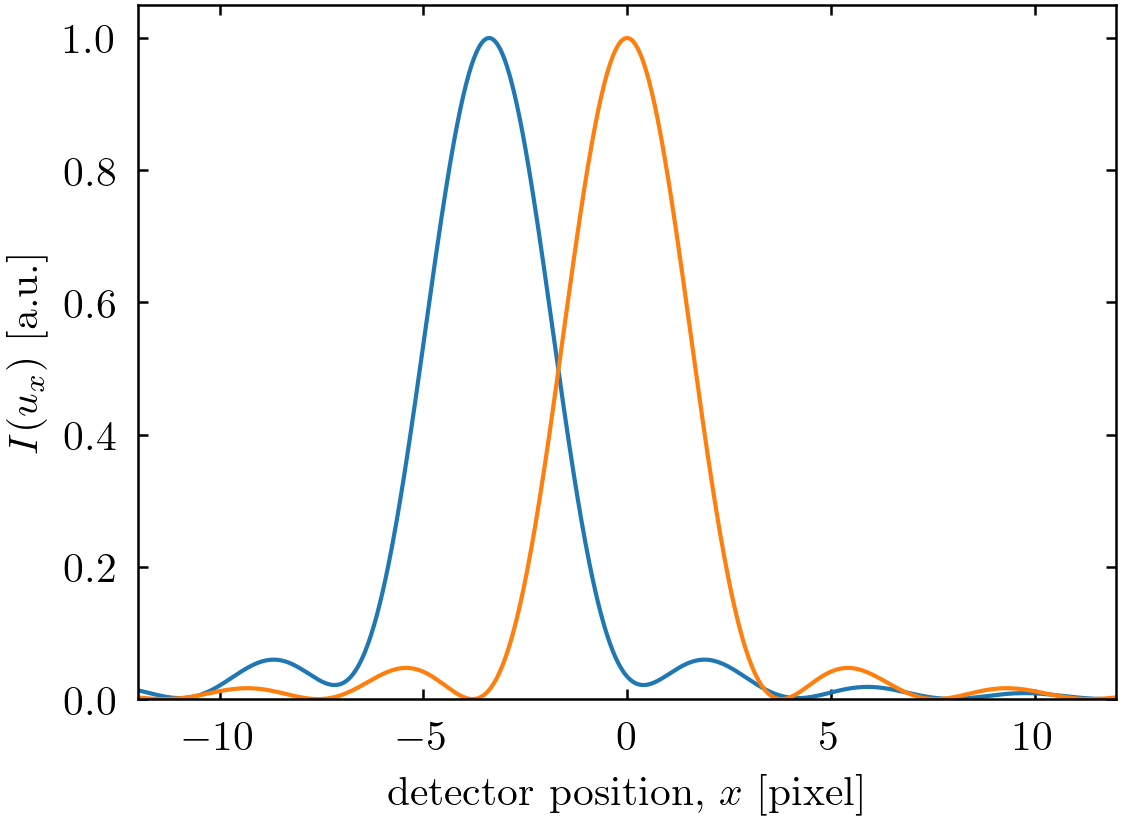}
  \caption{Predicted normalised intensity profile transverse to a diffraction spike, $I(u_x)$, from equation\,\ref{eq:diff_intensity}, for the case of zero WFE (orange curve) and after introduction of defocus, quadratic WFE (blue curve).
    \label{fig:fresnel}}
  \end{centering}
\end{figure}

For a pure defocus, the pupil-plane phase from \cref{eq:fraunhofer_general} is a radially symmetric, quadratic function dependent on radius $r$ measured from the pupil phase centre \citep{goodman_introduction_1996} and can be written as
\begin{equation}
  \phi(r) = \frac{2\pi d}{\lambda}\left(\frac{2 r}{D}\right)^2 \label{eq:defocus_phase}
\end{equation}
at wavelength $\lambda$, for $r \le D/2$,  where $D$ is the pupil diameter and $d$ is the optical-path difference at the outer radius of the pupil. 

A diffraction spike shift would arise for any orientation of the spider struts, as this effect only requires a phase gradient in the pupil plane along each strut.
To parametrise the phase shift in the \Euclid pupil, we employ the coordinate system shown in Fig.\,\ref{fig:schematic}.
In this basis, \cref{eq:defocus_phase} becomes
\begin{equation}
    \phi(x, y) = \frac{8\pi d}{\lambda}\frac{([y_0+y]^2+x^2)}{D^2} = \gamma \left([k_{y0}+k_y]^2 + k_x^2\right) \label{eq:euclid_defocus_phase} \, ,
\end{equation}
where $y_0$ is the perpendicular distance from the major axis of each strut to the phase centre, $k_{y0}=2\pi y_0/\lambda$ and for convenience we define $\gamma \equiv 2\lambda d/\pi D^2$.
The 2D inverse Fourier transform is then
\begin{equation}
  f(u_x,u_y) = \int_{{k_{x1}}}^{{k_{x2}}} {\rm e}^{{\rm i}(k_xu_x + \gamma k_x^2)}\,{\rm d}k_x
  \int_{{k_{y1}}}^{{k_{y2}}} {\rm e}^{{\rm i}(k_yu_y + \gamma (k_{y0}^2 + k_y^2))}\,{\rm d}k_y \, ,
\end{equation}
between wavenumber limits determined by the geometry of the strut.
We may separate the diffraction integral into orthogonal components: the
resulting diffraction spike is straight in angular coordinates, for the case of a pure defocus wavefront error.

The Fourier transform on the $k_x$-axis yields the transverse profile of the diffraction spike,
\begin{equation}
    f(u_x) = \left(\frac{\pi}{2\gamma}\right)^{1/2}{\rm e}^{-{\rm i}{u_x^2}/(4\gamma)}\int_{{\Phi_1}}^{{\Phi_2}} {\rm e}^{{\rm i}\pi{\Phi^2}/2} {\rm d}\Phi \, , \label{eq:fr_int}
\end{equation}
for $\gamma \ne 0$,
where we complete the square by defining $\Phi = (2/\pi)^{1/2}(\gamma^{\frac{1}{2}}k_x + \gamma^{-\frac{1}{2}}u_x/2)$
and where the integral is evaluated between limits in wavenumber $k_{x1}, k_{x2}$, or between limits $\Phi_1$, $\Phi_2$.
For $\gamma=0$ the integral evaluates to the usual sinc expression, but otherwise is a Fresnel integral with solution
\begin{align}
  f(u_x) =  \left(\frac{\pi}{2\gamma}\right)^{{1}/{2}} {\rm e}^{-{\rm i}{{u_x^2}/({4\gamma})}}
   \{
    &\left[\mathcal{F_C}\left(\Phi_2\right)-\mathcal{F_C}\left(\Phi_1\right)\right] \nonumber \\
    + \mathrm{i}
    &\left[\mathcal{F_S}\left(\Phi_2\right)-\mathcal{F_S}\left(\Phi_1\right)\right]
    \}
    \, , \label{eq:fresnel_integrals}
\end{align}
where $\mathcal{F_C}$, $\mathcal{F_S}$ are the Fresnel integrals \citep{born_principles_1999}.
The diffracted intensity at large radii, far from the core of the PSF, is then given by
\begin{equation}
I(u_x) \rightarrow | f(u_x) |^2 \, .
\label{eq:diff_intensity}
\end{equation}

The Fresnel integrals are odd functions and hence $I(u_x)$ is symmetric if we introduce a coordinate shift in direction cosine, $u_x' = u_x - u_0$,
such that $\Phi'_2 = -\Phi'_1$, from which \linebreak $u_0 = -\gamma\,(k_{x1}+k_{x2})=-2\,\gamma\,k_{x_\mathrm{c}}$, i.e. the diffraction intensity pattern
appears to shift laterally by $u_0$.  Given that the phase $\phi(k_x) = \gamma k_x^2$, we see that the direction cosine shift,
$u_0$, is given by the mean phase gradient along the strut,
\begin{equation}
  u_0 = -\left\langle {\rm d}\phi /{\rm d}k_x \right\rangle  \, . \label{eq:result}
\end{equation}
The lateral displacement of diffraction spikes in the focal plane is given by the product of $u_0$ and the telescope effective focal length.  
The intensity profile across a diffraction spike is shown in Fig.\,\ref{fig:fresnel}.  It can be seen that, although the transverse profile does vary slightly when defocus WFE is introduced, the dominant effect is a lateral shift.

Although evaluated here for the case of a pure defocus WFE,
and in reality other WFEs exist, we expect in general that the diffraction spikes should
shift laterally by a displacement dependent on their mean radial phase gradients.  Some other WFE modes also result in such phase gradients and hence diffraction spike shifts, including spherical aberration and trefoil (the latter because of the spin-3 symmetry of that mode). 

\begin{figure}
    \centering
    \includegraphics[width=1\linewidth]{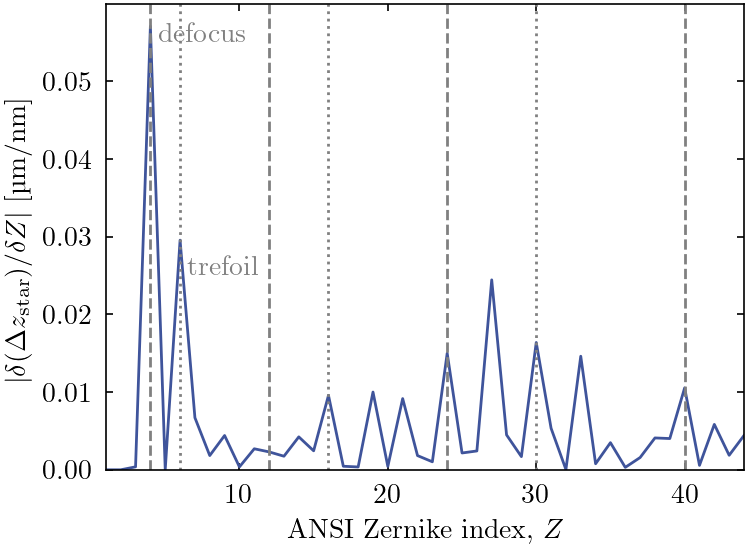}
    \caption{
    Sensitivity of the \zstar{} defocus estimate on the individual amplitude of Zernike WFE polynomials, designated by their ANSI index, assuming a linear relation. Radially symmetric wavefront errors are highlighted with dashed lines and (higher-order) trefoil modes with dotted lines. The sensitivity is averaged over a 3$\times$3 grid across the FoV. The highest sensitivities are to variations in the defocus $Z_2^0$, trefoil $Z_3^{-3}$, and $Z_6^6$ Zernike modes.
    }
    \label{fig:M2Z sensitivity}
\end{figure}

We tested the sensitivity to different WFE modes, as predicted from the \Euclid PSF model (Euclid Collaboration: Miller et al. 2026, in prep.)
% \citep{DR1-TP018} -> exchange above
by perturbing individual Zernike modes by $\pm10\,\si{\nano\meter}$ and linearly relating those perturbations to the corresponding change in measured $z_\mathrm{star}$. 
We averaged the sensitivity across the FoV in a 3$\times$3 grid. 
The resulting differential change per Zernike mode $Z$ perturbation is shown in \cref{fig:M2Z sensitivity}. 
As expected, \zstar{} is most sensitive to the actual defocus WFE ($Z$=4), although some higher order WFE influence our measure as well. However, during nominal survey, the defocus varies by an order of magnitude more than other WFE \citep{E-Anselmi}, consolidating that with this method we expect to trace actual defocus to good approximation. A method for correcting the defocus estimates for the sensitivity to trefoil WFE variations is presented by Euclid Collaboration: Whittam et al. (2026, in prep.),
% \citet{DR1-TP021}
where it is shown that this only has a small effect on the conclusions from the diffraction spike measurement post second ice-decontamination procedure.

Note that the lateral shift of \cref{eq:result}
is achromatic, for the case of WFE, $d$, introduced by an optical-path difference (\Cref{eq:defocus_phase}).
In the broadband \Euclid PSF, the lateral shift is the same at all wavelengths in the case of an
M2 defocus, leading to a clearly measurable effect.
Other chromatic sources of WFE, especially arising from the phase error of the
dichroic dielectric coating, would produce different, chromatic lateral shifts in the diffraction spikes,
causing some chromaticity and blurring of the shift, albeit diluted by the small wavelength ranges
over which the coating phase error is strongly chromatic.
This is the likely cause for the detected wavelength dependency displayed in \cref{fig:defocus_star_properties}.

\section{\label{apdx:non_uniformity_of_defocus_change}Field of view uniformity of defocus change}
\begin{figure*}
    \centering
    \includegraphics[width=1\linewidth]{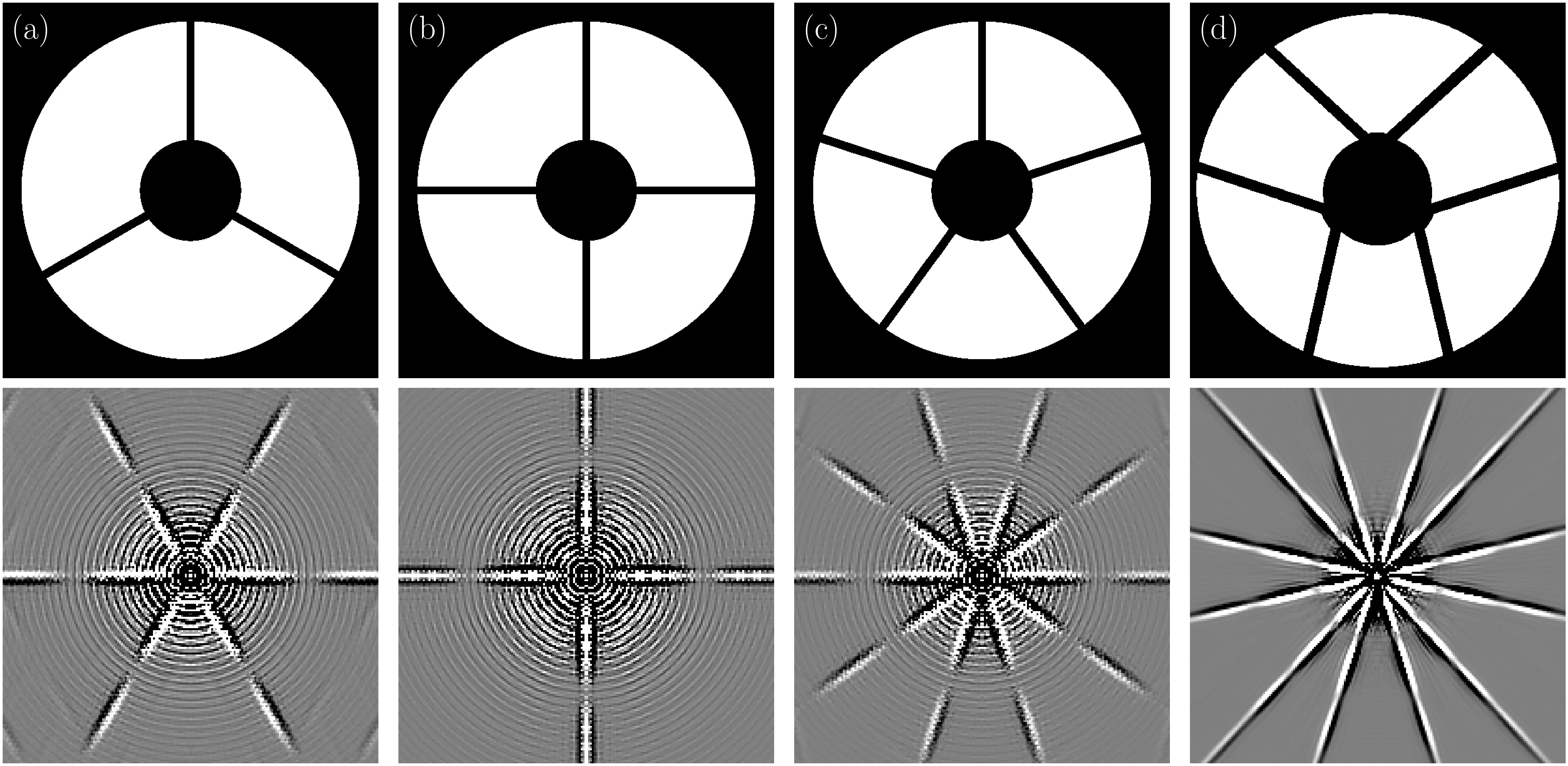}
    \renewcommand{\thefigure}{C.\arabic{figure}}
    \caption{
    \emph{Top}: Entrance pupil of an arbitrary diffraction limited telescope with three, four, and five spider arms (a, b, c, respectively), and of the \emph{Roman} space telescope (d). \emph{Bottom}: Corresponding difference between an in-focus PSF, and a 0.3 wavelength defocused PSF. Apart from a growing in size of the PSF core, the coherent diffraction spike shift for non-mirror-symmetric spider configuration is clearly visible.
}
    \label{fig:psf_pupil_diff}
\end{figure*}
\renewcommand{\thefigure}{\thesection.\arabic{figure}}% restore normal behavior
% \begin{figure}[ht]
%     \centering
%     \includegraphics[width=1\linewidth]{figures/polynomial_coefficients_in_defocus_bins_order1.png}
%     \caption{
%     Dependency of the fitted polynomial FoV coefficients on the total inferred defocus of the \Euclid VIS image.
%     The $c_i$ are defined by $\Delta z_\mathrm{star}=c_0+c_1\,y_\mathrm{FPA} + c_2\,x_\mathrm{FPA}$, with $\Delta z_\mathrm{star}$ being the defocus of individual stars as a function of position on the focal plane array ($x_\mathrm{FPA},\,y_\mathrm{FPA}$).
%     To highlight deviations from a constant offset, all $\Delta z_\mathrm{star}$ within a given $\Delta z_\mathrm{obs}$ bin are collected, binned in a 3$\times$3 pixel grid per CCD quadrant and subtracted by the median pixel values of the $[-0.61,\;-0.57]\,\micron$ before being fitted. Shaded regions depict the fitting error of the respective coefficient. \textcolor{red}{RENAME X,Y-LABELS}
%     }
%     \label{fig:binning_obs_in_defocus}
% \end{figure}

A deviation from a constant offset across the FoV can have two underlying causes:
\begin{enumerate}
    \item Non-trivial shifts of optical elements that result in a non-uniform deviation of the image plane from the focal point;
    \item A temporal evolution of the baseline FoV behaviour originating from changes in the general state of the telescope, be it sudden or creeping.
\end{enumerate}
We refer to \cref{sc:Results-deicing-influence} for the latter and focus here on the former. 

The deviations from the baseline $\Delta z_\mathrm{obs}$ apparent in \cref{fig:DefocusFullSurvey} give us the opportunity to study the non-uniform changes of $\Delta z_\mathrm{star}$ across the focal plane. 
The \Euclid spacecraft underwent two ice decontamination campaigns triggered by loss in throughput for light at the blue end of the VIS spectrum; the last one happened in June 2024 \citep{euclid_march_deicing, euclid_june_deicing}.
This procedure likely caused a significant change in underlying baseline defocus across the FoV (see \cref{sc:Results-deicing-influence}), which is why we restrict the analysis here to observations done from July 2024 onwards.
We bin the individual exposures in twenty equally-spaced $\Delta z_\mathrm{obs}$ bins between $-0.85\,\micron$ and $-0.25\,\micron$. 
% Information about the individual bins can be found in \cref{tab:binning_in_defocus_bin_summary}.
We collect the defocus of all stars falling within each specific bin onto a single image, take the median $\Delta z_\mathrm{star}$ per 2$\times$2 grid per CCD-quadrant and subtract the baseline FoV defocus shown in \cref{fig:fov_template_fiducial}.

To quantify the residual changes, we fit the second-order polynomial in Chebyshev base from \cref{eq:2d_polynomial_chebyshev} to the \zstar{} per defocus bin in \zobs{}.
We also tested a third-order polynomial fit, which affects the fit residual $\sigma_\mathrm{rms}(\Delta z_\mathrm{star})$ by less than 1\% in all \zobs{} bins, hence we restrict ourselves to the second-order polynomial.
After shifting all $z_\mathrm{obs}\rightarrow z_\mathrm{obs}-b_\mathrm{ref}$ where $b_\mathrm{ref}$ is the median defocus of the stable time frame from \cref{eq:diffspike_correction}, we fit a linear function to the individual $c_{ij}$,\linebreak such that $c_{ij}(\Delta z_\mathrm{obs})=a\,\Delta z_\mathrm{obs}+c_{ij,0}$.

If the assumption that the defocus changes uniformly across the FoV is correct, then any change of \zobs{} should be fully characterised by the constant offset $c_{00}$.
In this case, the slope $a$ would be unity for coefficient $c_{00}$ and zero for all other.
Furthermore, we expect all intersects $c_{ij,0}$ to coincide with zero if the FoV dependency of the reference period is representative of the FoV dependency across the full survey timeline post second ice decontamination.
\begin{table}[ht]
\caption{
Linear relation between second-order Chebyshev polynomial fit coefficients to the FoV difference per $\Delta z_\mathrm{obs}$ bin to the reference, along with the fit uncertainty. 
}
\label{tab:polynomial_fit_defocus_bins}
% \resizebox{0.5\textwidth}{!}{%
\begin{tabular}{l
                S[separate-uncertainty, table-format=1.5(2)]
                S[separate-uncertainty, table-format=1.5(2.1)]}
\hline \hline
Coefficient & {Slope $a$\,[\micron\,\micron$^{-1}$]} & {Intercept\,$c_{ij,0}$\,[\micron]} \\
\hline
$c_{00}$ & 0.9633 \pm 0.0022 & 0.00034 \pm 0.00017 \\
$c_{10}$ & 0.0138 \pm 0.0031 & 0.00157 \pm 0.00024 \\
$c_{01}$ & 0.0021 \pm 0.0031 & 0.00069 \pm 0.00024 \\
$c_{20}$ & 0.0136 \pm 0.0030 & -0.00100 \pm 0.00024 \\
$c_{11}$ & -0.0028 \pm 0.0053 & 0.00124 \pm 0.00041 \\
$c_{02}$ & -0.0055 \pm 0.0029 & 0.00194 \pm 0.00023 \\
\hline
\end{tabular}
% }
\tablefoot{The coefficients are labelled according to \cref{eq:2d_polynomial_chebyshev}. 
The first and last bin are excluded in the fit due to limited number of exposures.}
\end{table}

The results are displayed in \cref{tab:polynomial_fit_defocus_bins}.
The constant offset $c_{00}$ indeed grows linearly with $\Delta z_\mathrm{obs}$, although the deviation from unity is significant.
Additionally, the $x$-dependent coefficients $c_{10}$ and $c_{20}$ do not agree with zero.
Therefore, the defocus WFE changes with a slight tilt during nominal survey operations. 
This is not surprising if the defocus is caused by mechanical deformations due to changes in temperature, as we argue in Euclid Collaboration: Whittam et al. (2026, in prep.).
% \citet{DR1-TP021} -> exchange above
Still, the largest variations in \zobs{} since July 2024 shown in \cref{fig:DefocusFullSurvey} are on the order of $0.6\,\micron$.
By assuming a constant offset, we effectively absorb the FoV-dependent coefficients as a noise contribution into $c_{00}$ due to finite, discrete sampling of PSFs across the focal plane.
However, the additional RMS deviation around the median with a $\delta(\Delta z_\mathrm{obs})=0.6\,\micron$ are $\sigma_\mathrm{RMS}(\Delta z_\mathrm{star})=0.0076^{+0.0016}_{-0.0011}\,\micron$ and, as such, negligible in comparison to the defocus uncertainty per star from the spike detection method. 
We can therefore summarise the relative change in defocus into a constant offset across the FoV with marginal loss in accuracy for single exposures.

Additionally, the intercepts show deviations from zero. 
Therefore, the FoV variation from exposures binned since July 2024 is detectably different from the FoV variation binned over the full survey timeline.
In practice, however, the non-uniform $c_{ij,0}$ from \cref{tab:polynomial_fit_defocus_bins} induce an RMS around the median of $1.75\times10^{-3}\,\micron$ which is subdominant in comparison to the statistical uncertainty of our defocus inference method per star.
Therefore, within our statistical uncertainty, the conversion from \sqrtADS{} to $\Delta z$ on the individual star level with the (slope of the) calibrated relation \cref{eq:diffspike_to_defocus} is valid.

In this appendix, we showed that the defocus offsets are, to good approximation, homogeneous across the FoV. We adopt a methodology similar to that described in \cref{sc:Results-Temporal Evolution of FoV Template Uniformity}. As shown in \cref{tab:polynomial_fit_temporal_evolution}, the baseline defocus distribution across the FoV exhibits no significant temporal evolution. Similarly, \cref{tab:polynomial_fit_defocus_bins} indicates that variations in the median defocus lead to only negligible changes in the baseline distribution.

Taken together, we argue that the defocus response to telescope perturbations during nominal survey operations is uniform across the FoV, strengthening the conclusion that the shifts observed during the ice-decontamination campaigns reflect substantial modifications of the underlying PSF, while also underscoring the remarkable stability of the spacecraft ever since.

\section{\label{apdx:pupil_configurations}Diffraction spike shifts for different entrance pupil configurations}

In this section, we motivate the influence of defocus WFE on the PSF originating from different telescope entrance pupil configurations. 
The images are generated with the \texttt{GalSim} simulation suite\footnote{\url{https://github.com/GalSim-developers/GalSim}}, assuming a flat spectral energy distribution in photon count.
We present a generic telescope with 3, 4, and 5 spider arms, a pixel scale of $\ang{;;0.1}/$pixel, flat bandpass between 550--900\,nm, a fiducial wavelength of 700\,nm, and a primary mirror diameter of 1.2\,m. 
We additionally use the \texttt{GalSim}-\emph{Roman} package to create images within the F129 bandpass with a fiducial wavelength of 1293\,nm.

The influence of defocus on each corresponding PSF is visualised in \cref{fig:psf_pupil_diff}. The top row shows the entrance pupil and the bottom row the difference between an in-focus PSF, and one that has been defocused by 0.3 wavelengths. 
We can distinguish two types of pupil vane configuration: vanes that have no identical counterpart on the opposite end of the pupil, and those that have one.\footnote{One can refer to this as non-mirror symmetric, but has to be careful with the semantics. The mirroring axis has to be drawn perpendicular to individual spike extensions, i.e. along the $y$-axis in \cref{fig:schematic}.}
Pupils (a), (c), and (d) belong to the first category and the coherent shift in diffraction spike is clearly visible. However, for configurations of the second category like (b), the shift of individual spikes becomes obscured since they overlap. 
In particular for small defoci a detection of the spikes with brightness moments becomes impossible due to the blending. 
However, one could imagine using a different metric for defocus estimation (within a limited range) that quantifies the width of the overlapping diffraction spikes, although the direction of defocus cannot be uniquely determined.
Note that (d) is the entrance pupil for the \emph{Roman} space telescope. 
Here, the enclosed area between all spikes creates a complex polygon due to angled spider struts.
However, we can easily reduce the complexity of the problem by just selecting a subset of three diffraction spikes, allowing us to apply the same detection philosophy as for the \Euclid pupil.
In fact, we can build two independent realisations of the diffraction spike triangle by using two distinct sets of diffraction spike-triplets.

\setcounter{figure}{1}% so this becomes C.2
\begin{figure}[t]
    \centering
    \includegraphics[width=1\linewidth]{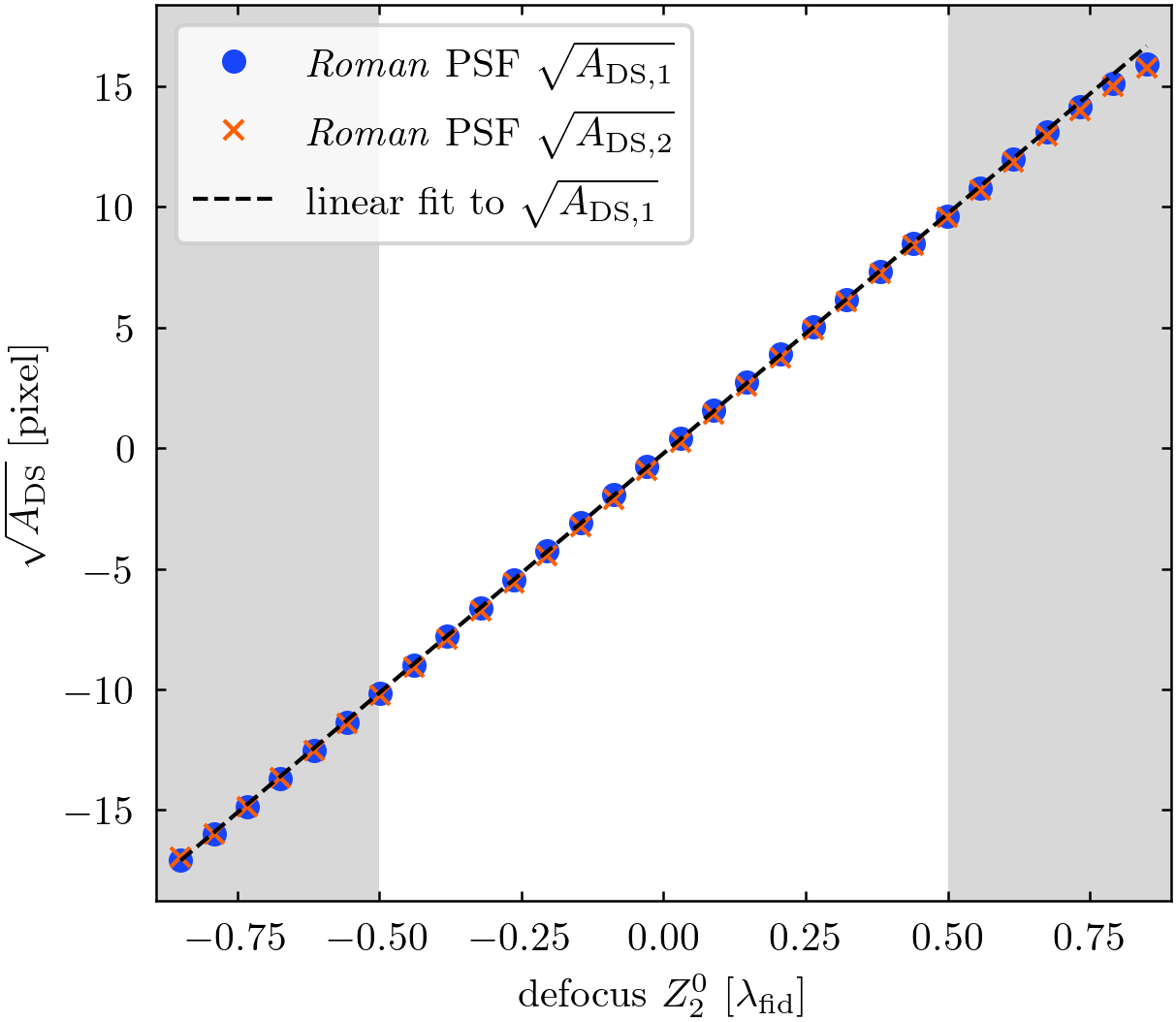}
    \caption{Linear relation between the enclosing diffraction spike area for two different spike-triplets $\sqrt{A_\mathrm{DS,1/2}}$ of the \emph{Roman} space telescope PSF, as a function of Zernike defocus mode amplitude in units of the fiducial 1293\,nm wavelength.
    The black line shows the linear fit for the $\sqrt{A_\mathrm{DS,1}}$ case, excluding defocus ranges within the grey bands.
}
    \label{fig:Roman_sqrtAds_vs_defocus}
\end{figure}

With the \texttt{GalSim}-\emph{Roman} package, we can verify linearity between defocus and \sqrtADS{} over a large perturbation range. 
We split the \emph{Roman} PSF diffraction spikes into two sets of (approximate, visually selected) angles:
\begin{align}
    \label{eq:Roman_diffspike_angles}
    \vec{\theta}_1&=(12.5\degree, 71.5\degree, 132.5\degree)\,,\nonumber\\
    \vec{\theta}_2&=(47\degree, 108.5\degree, 167.5\degree)\,.
\end{align}
We apply defocus by perturbing Zernike Noll mode 4 ($Z_2^0$) between $\pm0.85$ fiducial wavelengths.
Then, we measure $\sqrt{A_\mathrm{DS,1/2}}$, where the index corresponds to the triangle formed by either set of \cref{eq:Roman_diffspike_angles}.
The relation between $Z_2^0$ and $\sqrt{A_\mathrm{DS,1/2}}$ is shown in \cref{fig:Roman_sqrtAds_vs_defocus}.
Our choice of the angle sub-selection is purposefully done such that both enclosing triangles have a similar size. 
This apparently works well in the image simulations, and both $\sqrt{A_\mathrm{DS,1/2}}$ evolve linear with defocus.
Therefore, we expect our presented method to be easily applicable to the \emph{Roman} space telescope.

\end{appendix}

\label{LastPage}
\end{document}

%% file: authors.tex
%%%% Version Thursday 18th of June 2026 07:12:20 AM UT
%%%% Assumes the new A&A style file from Oct 2025 or later
%%%% Please do not edit the author list -- contact ECEB Bureau for changes
\newcommand{\orcid}[1]{} %% if already defined in aa.cls: comment, or use renewcommand			   
\author{D.~Neumann\orcid{0009-0006-3926-6756}\thanks{\email{dneumann@strw.leidenuniv.nl}}\inst{\ref{aff1}}
\and L.~Miller\orcid{0000-0002-3376-6200}\inst{\ref{aff2}}
\and H.~Hoekstra\orcid{0000-0002-0641-3231}\inst{\ref{aff1}}
\and K.~Kuijken\orcid{0000-0002-3827-0175}\inst{\ref{aff1}}
\and I.~H.~Whittam\orcid{0000-0003-2265-5983}\inst{\ref{aff2},\ref{aff3}}
\and N.~E.~Chisari\orcid{0000-0003-4221-6718}\inst{\ref{aff4},\ref{aff1}}
\and R.~Nakajima\orcid{0009-0009-1213-7040}\inst{\ref{aff5}}
\and B.~Altieri\orcid{0000-0003-3936-0284}\inst{\ref{aff6}}
\and A.~Amara\inst{\ref{aff7}}
\and S.~Andreon\orcid{0000-0002-2041-8784}\inst{\ref{aff8}}
\and N.~Auricchio\orcid{0000-0003-4444-8651}\inst{\ref{aff9}}
\and C.~Baccigalupi\orcid{0000-0002-8211-1630}\inst{\ref{aff10},\ref{aff11},\ref{aff12},\ref{aff13}}
\and M.~Baldi\orcid{0000-0003-4145-1943}\inst{\ref{aff14},\ref{aff9},\ref{aff15}}
\and S.~Bardelli\orcid{0000-0002-8900-0298}\inst{\ref{aff9}}
\and A.~Basset\inst{\ref{aff16}}
\and P.~Battaglia\orcid{0000-0002-7337-5909}\inst{\ref{aff9}}
\and A.~Biviano\orcid{0000-0002-0857-0732}\inst{\ref{aff11},\ref{aff10}}
\and E.~Branchini\orcid{0000-0002-0808-6908}\inst{\ref{aff17},\ref{aff18},\ref{aff8}}
\and M.~Brescia\orcid{0000-0001-9506-5680}\inst{\ref{aff19},\ref{aff20}}
\and S.~Camera\orcid{0000-0003-3399-3574}\inst{\ref{aff21},\ref{aff22},\ref{aff23}}
\and G.~Ca\~nas-Herrera\orcid{0000-0003-2796-2149}\inst{\ref{aff1}}
\and V.~Capobianco\orcid{0000-0002-3309-7692}\inst{\ref{aff23}}
\and C.~Carbone\orcid{0000-0003-0125-3563}\inst{\ref{aff24}}
\and J.~Carretero\orcid{0000-0002-3130-0204}\inst{\ref{aff25},\ref{aff26}}
\and M.~Castellano\orcid{0000-0001-9875-8263}\inst{\ref{aff27}}
\and G.~Castignani\orcid{0000-0001-6831-0687}\inst{\ref{aff9}}
\and S.~Cavuoti\orcid{0000-0002-3787-4196}\inst{\ref{aff20},\ref{aff28}}
\and K.~C.~Chambers\orcid{0000-0001-6965-7789}\inst{\ref{aff29}}
\and A.~Cimatti\inst{\ref{aff30}}
\and C.~Colodro-Conde\inst{\ref{aff31}}
\and G.~Congedo\orcid{0000-0003-2508-0046}\inst{\ref{aff32}}
\and C.~J.~Conselice\orcid{0000-0003-1949-7638}\inst{\ref{aff33}}
\and L.~Conversi\orcid{0000-0002-6710-8476}\inst{\ref{aff34},\ref{aff6}}
\and Y.~Copin\orcid{0000-0002-5317-7518}\inst{\ref{aff35}}
\and F.~Courbin\orcid{0000-0003-0758-6510}\inst{\ref{aff36},\ref{aff37},\ref{aff38}}
\and H.~M.~Courtois\orcid{0000-0003-0509-1776}\inst{\ref{aff39}}
\and M.~Cropper\orcid{0000-0003-4571-9468}\inst{\ref{aff40}}
\and H.~Degaudenzi\orcid{0000-0002-5887-6799}\inst{\ref{aff41}}
\and G.~De~Lucia\orcid{0000-0002-6220-9104}\inst{\ref{aff11}}
\and H.~Dole\orcid{0000-0002-9767-3839}\inst{\ref{aff42}}
\and F.~Dubath\orcid{0000-0002-6533-2810}\inst{\ref{aff41}}
\and X.~Dupac\inst{\ref{aff6}}
\and M.~Farina\orcid{0000-0002-3089-7846}\inst{\ref{aff43}}
\and R.~Farinelli\inst{\ref{aff9}}
\and S.~Farrens\orcid{0000-0002-9594-9387}\inst{\ref{aff44}}
\and S.~Ferriol\inst{\ref{aff35}}
\and S.~Fotopoulou\orcid{0000-0002-9686-254X}\inst{\ref{aff45}}
\and N.~Fourmanoit\orcid{0009-0005-6816-6925}\inst{\ref{aff46}}
\and M.~Frailis\orcid{0000-0002-7400-2135}\inst{\ref{aff11}}
\and E.~Franceschi\orcid{0000-0002-0585-6591}\inst{\ref{aff9}}
\and M.~Fumana\orcid{0000-0001-6787-5950}\inst{\ref{aff24}}
\and S.~Galeotta\orcid{0000-0002-3748-5115}\inst{\ref{aff11}}
\and K.~George\orcid{0000-0002-1734-8455}\inst{\ref{aff47}}
\and B.~Gillis\orcid{0000-0002-4478-1270}\inst{\ref{aff32}}
\and C.~Giocoli\orcid{0000-0002-9590-7961}\inst{\ref{aff9},\ref{aff15}}
\and P.~G\'omez-Alvarez\orcid{0000-0002-8594-5358}\inst{\ref{aff48},\ref{aff6}}
\and J.~Gracia-Carpio\orcid{0000-0003-4689-3134}\inst{\ref{aff49}}
\and A.~Grazian\orcid{0000-0002-5688-0663}\inst{\ref{aff50}}
\and F.~Grupp\inst{\ref{aff49},\ref{aff51}}
\and S.~V.~H.~Haugan\orcid{0000-0001-9648-7260}\inst{\ref{aff52}}
\and W.~Holmes\orcid{0009-0007-8554-4646}\inst{\ref{aff53}}
\and F.~Hormuth\inst{\ref{aff54}}
\and A.~Hornstrup\orcid{0000-0002-3363-0936}\inst{\ref{aff55},\ref{aff56}}
\and K.~Jahnke\orcid{0000-0003-3804-2137}\inst{\ref{aff57}}
\and M.~Jhabvala\inst{\ref{aff58}}
\and B.~Joachimi\orcid{0000-0001-7494-1303}\inst{\ref{aff59}}
\and S.~Kermiche\orcid{0000-0002-0302-5735}\inst{\ref{aff46}}
\and A.~Kiessling\orcid{0000-0002-2590-1273}\inst{\ref{aff53}}
\and M.~Kilbinger\orcid{0000-0001-9513-7138}\inst{\ref{aff44}}
\and R.~Kohley\inst{\ref{aff6}}
\and B.~Kubik\orcid{0009-0006-5823-4880}\inst{\ref{aff35}}
\and M.~Kunz\orcid{0000-0002-3052-7394}\inst{\ref{aff60}}
\and H.~Kurki-Suonio\orcid{0000-0002-4618-3063}\inst{\ref{aff61},\ref{aff62}}
\and R.~Laureijs\inst{\ref{aff63}}
\and A.~M.~C.~Le~Brun\orcid{0000-0002-0936-4594}\inst{\ref{aff64}}
\and S.~Ligori\orcid{0000-0003-4172-4606}\inst{\ref{aff23}}
\and P.~B.~Lilje\orcid{0000-0003-4324-7794}\inst{\ref{aff52}}
\and V.~Lindholm\orcid{0000-0003-2317-5471}\inst{\ref{aff61},\ref{aff62}}
\and I.~Lloro\orcid{0000-0001-5966-1434}\inst{\ref{aff65}}
\and G.~Mainetti\orcid{0000-0003-2384-2377}\inst{\ref{aff66}}
\and O.~Mansutti\orcid{0000-0001-5758-4658}\inst{\ref{aff11}}
\and O.~Marggraf\orcid{0000-0001-7242-3852}\inst{\ref{aff5}}
\and M.~Martinelli\orcid{0000-0002-6943-7732}\inst{\ref{aff27},\ref{aff67}}
\and N.~Martinet\orcid{0000-0003-2786-7790}\inst{\ref{aff68}}
\and F.~Marulli\orcid{0000-0002-8850-0303}\inst{\ref{aff69},\ref{aff9},\ref{aff15}}
\and R.~J.~Massey\orcid{0000-0002-6085-3780}\inst{\ref{aff70}}
\and E.~Medinaceli\orcid{0000-0002-4040-7783}\inst{\ref{aff9}}
\and S.~Mei\orcid{0000-0002-2849-559X}\inst{\ref{aff71},\ref{aff72}}
\and M.~Meneghetti\orcid{0000-0003-1225-7084}\inst{\ref{aff9},\ref{aff15}}
\and E.~Merlin\orcid{0000-0001-6870-8900}\inst{\ref{aff27}}
\and G.~Meylan\orcid{0000-0001-6503-0209}\inst{\ref{aff73}}
\and A.~Mora\orcid{0000-0002-1922-8529}\inst{\ref{aff74}}
\and M.~Moresco\orcid{0000-0002-7616-7136}\inst{\ref{aff69},\ref{aff9}}
\and C.~Moretti\orcid{0000-0003-3314-8936}\inst{\ref{aff11},\ref{aff10},\ref{aff12}}
\and L.~Moscardini\orcid{0000-0002-3473-6716}\inst{\ref{aff69},\ref{aff9},\ref{aff15}}
\and C.~Neissner\orcid{0000-0001-8524-4968}\inst{\ref{aff75},\ref{aff26}}
\and R.~C.~Nichol\orcid{0000-0003-0939-6518}\inst{\ref{aff7}}
\and S.-M.~Niemi\orcid{0009-0005-0247-0086}\inst{\ref{aff76}}
\and C.~Padilla\orcid{0000-0001-7951-0166}\inst{\ref{aff75}}
\and S.~Paltani\orcid{0000-0002-8108-9179}\inst{\ref{aff41}}
\and F.~Pasian\orcid{0000-0002-4869-3227}\inst{\ref{aff11}}
\and K.~Pedersen\inst{\ref{aff77}}
\and W.~J.~Percival\orcid{0000-0002-0644-5727}\inst{\ref{aff78},\ref{aff79},\ref{aff80}}
\and V.~Pettorino\orcid{0000-0002-4203-9320}\inst{\ref{aff76}}
\and A.~Pezzotta\orcid{0000-0003-0726-2268}\inst{\ref{aff8}}
\and S.~Pires\orcid{0000-0002-0249-2104}\inst{\ref{aff44}}
\and G.~Polenta\orcid{0000-0003-4067-9196}\inst{\ref{aff81}}
\and M.~Poncet\inst{\ref{aff16}}
\and L.~A.~Popa\inst{\ref{aff82}}
\and G.~D.~Racca\orcid{0000-0002-9883-8981}\inst{\ref{aff1},\ref{aff76}}
\and F.~Raison\orcid{0000-0002-7819-6918}\inst{\ref{aff49}}
\and A.~Renzi\orcid{0000-0001-9856-1970}\inst{\ref{aff83},\ref{aff84},\ref{aff9}}
\and J.~Rhodes\orcid{0000-0002-4485-8549}\inst{\ref{aff53}}
\and G.~Riccio\inst{\ref{aff20}}
\and E.~Romelli\orcid{0000-0003-3069-9222}\inst{\ref{aff11}}
\and M.~Roncarelli\orcid{0000-0001-9587-7822}\inst{\ref{aff9}}
\and R.~Saglia\orcid{0000-0003-0378-7032}\inst{\ref{aff51},\ref{aff49}}
\and Z.~Sakr\orcid{0000-0002-4823-3757}\inst{\ref{aff85},\ref{aff86},\ref{aff87}}
\and D.~Sapone\orcid{0000-0001-7089-4503}\inst{\ref{aff88}}
\and B.~Sartoris\orcid{0000-0003-1337-5269}\inst{\ref{aff51},\ref{aff11}}
\and M.~Schirmer\orcid{0000-0003-2568-9994}\inst{\ref{aff57}}
\and P.~Schneider\orcid{0000-0001-8561-2679}\inst{\ref{aff5}}
\and T.~Schrabback\orcid{0000-0002-6987-7834}\inst{\ref{aff89}}
\and A.~Secroun\orcid{0000-0003-0505-3710}\inst{\ref{aff46}}
\and E.~Sihvola\orcid{0000-0003-1804-7715}\inst{\ref{aff90}}
\and P.~Simon\inst{\ref{aff5}}
\and C.~Sirignano\orcid{0000-0002-0995-7146}\inst{\ref{aff83},\ref{aff84}}
\and G.~Sirri\orcid{0000-0003-2626-2853}\inst{\ref{aff15}}
\and A.~Spurio~Mancini\orcid{0000-0001-5698-0990}\inst{\ref{aff91}}
\and L.~Stanco\orcid{0000-0002-9706-5104}\inst{\ref{aff84}}
\and P.~Tallada-Cresp\'{i}\orcid{0000-0002-1336-8328}\inst{\ref{aff25},\ref{aff26}}
\and A.~N.~Taylor\inst{\ref{aff32}}
\and I.~Tereno\orcid{0000-0002-4537-6218}\inst{\ref{aff92},\ref{aff93}}
\and N.~Tessore\orcid{0000-0002-9696-7931}\inst{\ref{aff40}}
\and S.~Toft\orcid{0000-0003-3631-7176}\inst{\ref{aff94},\ref{aff95}}
\and R.~Toledo-Moreo\orcid{0000-0002-2997-4859}\inst{\ref{aff96},\ref{aff97}}
\and F.~Torradeflot\orcid{0000-0003-1160-1517}\inst{\ref{aff26},\ref{aff25}}
\and I.~Tutusaus\orcid{0000-0002-3199-0399}\inst{\ref{aff98},\ref{aff99},\ref{aff86}}
\and L.~Valenziano\orcid{0000-0002-1170-0104}\inst{\ref{aff9},\ref{aff100}}
\and J.~Valiviita\orcid{0000-0001-6225-3693}\inst{\ref{aff61},\ref{aff62}}
\and T.~Vassallo\orcid{0000-0001-6512-6358}\inst{\ref{aff11},\ref{aff47}}
\and A.~Veropalumbo\orcid{0000-0003-2387-1194}\inst{\ref{aff8},\ref{aff18},\ref{aff17}}
\and Y.~Wang\orcid{0000-0002-4749-2984}\inst{\ref{aff101}}
\and J.~Weller\orcid{0000-0002-8282-2010}\inst{\ref{aff51},\ref{aff49}}
\and G.~Zamorani\orcid{0000-0002-2318-301X}\inst{\ref{aff9}}
\and F.~M.~Zerbi\orcid{0000-0002-9996-973X}\inst{\ref{aff8}}
\and A.~Gregorio\orcid{0000-0003-4028-8785}\inst{\ref{aff102},\ref{aff11},\ref{aff12}}
\and A.~Loureiro\orcid{0000-0002-4371-0876}\inst{\ref{aff103},\ref{aff104}}
\and M.~Sereno\orcid{0000-0003-0302-0325}\inst{\ref{aff9},\ref{aff15}}}
										   
%%%% please do not edit the affiliation list -- contact ECEB Bureau for changes
\institute{Leiden Observatory, Leiden University, Einsteinweg 55, 2333 CC Leiden, The Netherlands\label{aff1}
\and
Department of Physics, Oxford University, Keble Road, Oxford OX1 3RH, UK\label{aff2}
\and
Department of Physics and Astronomy, University of the Western Cape, Bellville, Cape Town, 7535, South Africa\label{aff3}
\and
Institute for Theoretical Physics, Utrecht University, Princetonplein 5, 3584 CE Utrecht, The Netherlands\label{aff4}
\and
Universit\"at Bonn, Argelander-Institut f\"ur Astronomie, Auf dem H\"ugel 71, 53121 Bonn, Germany\label{aff5}
\and
ESAC/ESA, Camino Bajo del Castillo, s/n., Urb. Villafranca del Castillo, 28692 Villanueva de la Ca\~nada, Madrid, Spain\label{aff6}
\and
School of Mathematics and Physics, University of Surrey, Guildford, Surrey, GU2 7XH, UK\label{aff7}
\and
INAF-Osservatorio Astronomico di Brera, Via Brera 28, 20122 Milano, Italy\label{aff8}
\and
INAF-Osservatorio di Astrofisica e Scienza dello Spazio di Bologna, Via Piero Gobetti 93/3, 40129 Bologna, Italy\label{aff9}
\and
IFPU, Institute for Fundamental Physics of the Universe, via Beirut 2, 34151 Trieste, Italy\label{aff10}
\and
INAF-Osservatorio Astronomico di Trieste, Via G. B. Tiepolo 11, 34143 Trieste, Italy\label{aff11}
\and
INFN, Sezione di Trieste, Via Valerio 2, 34127 Trieste TS, Italy\label{aff12}
\and
SISSA, International School for Advanced Studies, Via Bonomea 265, 34136 Trieste TS, Italy\label{aff13}
\and
Dipartimento di Fisica e Astronomia, Universit\`a di Bologna, Via Gobetti 93/2, 40129 Bologna, Italy\label{aff14}
\and
INFN-Sezione di Bologna, Viale Berti Pichat 6/2, 40127 Bologna, Italy\label{aff15}
\and
Centre National d'Etudes Spatiales -- Centre spatial de Toulouse, 18 avenue Edouard Belin, 31401 Toulouse Cedex 9, France\label{aff16}
\and
Dipartimento di Fisica, Universit\`a di Genova, Via Dodecaneso 33, 16146, Genova, Italy\label{aff17}
\and
INFN-Sezione di Genova, Via Dodecaneso 33, 16146, Genova, Italy\label{aff18}
\and
Department of Physics "E. Pancini", University Federico II, Via Cinthia 6, 80126, Napoli, Italy\label{aff19}
\and
INAF-Osservatorio Astronomico di Capodimonte, Via Moiariello 16, 80131 Napoli, Italy\label{aff20}
\and
Dipartimento di Fisica, Universit\`a degli Studi di Torino, Via P. Giuria 1, 10125 Torino, Italy\label{aff21}
\and
INFN-Sezione di Torino, Via P. Giuria 1, 10125 Torino, Italy\label{aff22}
\and
INAF-Osservatorio Astrofisico di Torino, Via Osservatorio 20, 10025 Pino Torinese (TO), Italy\label{aff23}
\and
INAF-IASF Milano, Via Alfonso Corti 12, 20133 Milano, Italy\label{aff24}
\and
Centro de Investigaciones Energ\'eticas, Medioambientales y Tecnol\'ogicas (CIEMAT), Avenida Complutense 40, 28040 Madrid, Spain\label{aff25}
\and
Port d'Informaci\'{o} Cient\'{i}fica, Campus UAB, C. Albareda s/n, 08193 Bellaterra (Barcelona), Spain\label{aff26}
\and
INAF-Osservatorio Astronomico di Roma, Via Frascati 33, 00078 Monteporzio Catone, Italy\label{aff27}
\and
INFN section of Naples, Via Cinthia 6, 80126, Napoli, Italy\label{aff28}
\and
Institute for Astronomy, University of Hawaii, 2680 Woodlawn Drive, Honolulu, HI 96822, USA\label{aff29}
\and
Dipartimento di Fisica e Astronomia "Augusto Righi" - Alma Mater Studiorum Universit\`a di Bologna, Viale Berti Pichat 6/2, 40127 Bologna, Italy\label{aff30}
\and
Instituto de Astrof\'{\i}sica de Canarias, E-38205 La Laguna, Tenerife, Spain\label{aff31}
\and
Institute for Astronomy, University of Edinburgh, Royal Observatory, Blackford Hill, Edinburgh EH9 3HJ, UK\label{aff32}
\and
Jodrell Bank Centre for Astrophysics, Department of Physics and Astronomy, University of Manchester, Oxford Road, Manchester M13 9PL, UK\label{aff33}
\and
European Space Agency/ESRIN, Largo Galileo Galilei 1, 00044 Frascati, Roma, Italy\label{aff34}
\and
Universit\'e Claude Bernard Lyon 1, CNRS/IN2P3, IP2I Lyon, UMR 5822, Villeurbanne, F-69100, France\label{aff35}
\and
Institut de Ci\`{e}ncies del Cosmos (ICCUB), Universitat de Barcelona (IEEC-UB), Mart\'{i} i Franqu\`{e}s 1, 08028 Barcelona, Spain\label{aff36}
\and
Instituci\'o Catalana de Recerca i Estudis Avan\c{c}ats (ICREA), Passeig de Llu\'{\i}s Companys 23, 08010 Barcelona, Spain\label{aff37}
\and
Institut de Ciencies de l'Espai (IEEC-CSIC), Campus UAB, Carrer de Can Magrans, s/n Cerdanyola del Vall\'es, 08193 Barcelona, Spain\label{aff38}
\and
UCB Lyon 1, CNRS/IN2P3, IUF, IP2I Lyon, 4 rue Enrico Fermi, 69622 Villeurbanne, France\label{aff39}
\and
Mullard Space Science Laboratory, University College London, Holmbury St Mary, Dorking, Surrey RH5 6NT, UK\label{aff40}
\and
Department of Astronomy, University of Geneva, ch. d'Ecogia 16, 1290 Versoix, Switzerland\label{aff41}
\and
Universit\'e Paris-Saclay, CNRS, Institut d'astrophysique spatiale, 91405, Orsay, France\label{aff42}
\and
INAF-Istituto di Astrofisica e Planetologia Spaziali, via del Fosso del Cavaliere, 100, 00100 Roma, Italy\label{aff43}
\and
Universit\'e Paris-Saclay, Universit\'e Paris Cit\'e, CEA, CNRS, AIM, 91191, Gif-sur-Yvette, France\label{aff44}
\and
School of Physics, HH Wills Physics Laboratory, University of Bristol, Tyndall Avenue, Bristol, BS8 1TL, UK\label{aff45}
\and
Aix-Marseille Universit\'e, CNRS/IN2P3, CPPM, Marseille, France\label{aff46}
\and
University Observatory, LMU Faculty of Physics, Scheinerstr.~1, 81679 Munich, Germany\label{aff47}
\and
FRACTAL S.L.N.E., calle Tulip\'an 2, Portal 13 1A, 28231, Las Rozas de Madrid, Spain\label{aff48}
\and
Max Planck Institute for Extraterrestrial Physics, Giessenbachstr. 1, 85748 Garching, Germany\label{aff49}
\and
INAF-Osservatorio Astronomico di Padova, Via dell'Osservatorio 5, 35122 Padova, Italy\label{aff50}
\and
Universit\"ats-Sternwarte M\"unchen, Fakult\"at f\"ur Physik, Ludwig-Maximilians-Universit\"at M\"unchen, Scheinerstr.~1, 81679 M\"unchen, Germany\label{aff51}
\and
Institute of Theoretical Astrophysics, University of Oslo, P.O. Box 1029 Blindern, 0315 Oslo, Norway\label{aff52}
\and
Jet Propulsion Laboratory, California Institute of Technology, 4800 Oak Grove Drive, Pasadena, CA, 91109, USA\label{aff53}
\and
Felix Hormuth Engineering, Goethestr. 17, 69181 Leimen, Germany\label{aff54}
\and
Technical University of Denmark, Elektrovej 327, 2800 Kgs. Lyngby, Denmark\label{aff55}
\and
Cosmic Dawn Center (DAWN), Denmark\label{aff56}
\and
Max-Planck-Institut f\"ur Astronomie, K\"onigstuhl 17, 69117 Heidelberg, Germany\label{aff57}
\and
NASA Goddard Space Flight Center, Greenbelt, MD 20771, USA\label{aff58}
\and
Department of Physics and Astronomy, University College London, Gower Street, London WC1E 6BT, UK\label{aff59}
\and
Universit\'e de Gen\`eve, D\'epartement de Physique Th\'eorique and Centre for Astroparticle Physics, 24 quai Ernest-Ansermet, CH-1211 Gen\`eve 4, Switzerland\label{aff60}
\and
Department of Physics, P.O. Box 64, University of Helsinki, 00014 Helsinki, Finland\label{aff61}
\and
Helsinki Institute of Physics, Gustaf H{\"a}llstr{\"o}min katu 2, University of Helsinki, 00014 Helsinki, Finland\label{aff62}
\and
Kapteyn Astronomical Institute, University of Groningen, PO Box 800, 9700 AV Groningen, The Netherlands\label{aff63}
\and
Laboratoire d'etude de l'Univers et des phenomenes eXtremes, Observatoire de Paris, Universit\'e PSL, Sorbonne Universit\'e, CNRS, 92190 Meudon, France\label{aff64}
\and
SKAO, Jodrell Bank, Lower Withington, Macclesfield SK11 9FT, UK\label{aff65}
\and
Centre de Calcul de l'IN2P3/CNRS, 21 avenue Pierre de Coubertin 69627 Villeurbanne Cedex, France\label{aff66}
\and
INFN-Sezione di Roma, Piazzale Aldo Moro, 2 - c/o Dipartimento di Fisica, Edificio G. Marconi, 00185 Roma, Italy\label{aff67}
\and
Aix-Marseille Universit\'e, CNRS, CNES, LAM, Marseille, France\label{aff68}
\and
Dipartimento di Fisica e Astronomia "Augusto Righi" - Alma Mater Studiorum Universit\`a di Bologna, via Piero Gobetti 93/2, 40129 Bologna, Italy\label{aff69}
\and
Department of Physics, Institute for Computational Cosmology, Durham University, South Road, Durham, DH1 3LE, UK\label{aff70}
\and
Universit\'e Paris Cit\'e, CNRS, Astroparticule et Cosmologie, 75013 Paris, France\label{aff71}
\and
CNRS-UCB International Research Laboratory, Centre Pierre Bin\'etruy, IRL2007, CPB-IN2P3, Berkeley, USA\label{aff72}
\and
Institute of Physics, Laboratory of Astrophysics, Ecole Polytechnique F\'ed\'erale de Lausanne (EPFL), Observatoire de Sauverny, 1290 Versoix, Switzerland\label{aff73}
\and
Telespazio UK S.L. for European Space Agency (ESA), Camino bajo del Castillo, s/n, Urbanizacion Villafranca del Castillo, Villanueva de la Ca\~nada, 28692 Madrid, Spain\label{aff74}
\and
Institut de F\'{i}sica d'Altes Energies (IFAE), The Barcelona Institute of Science and Technology, Campus UAB, 08193 Bellaterra (Barcelona), Spain\label{aff75}
\and
European Space Agency/ESTEC, Keplerlaan 1, 2201 AZ Noordwijk, The Netherlands\label{aff76}
\and
DARK, Niels Bohr Institute, University of Copenhagen, Jagtvej 155, 2200 Copenhagen, Denmark\label{aff77}
\and
Waterloo Centre for Astrophysics, University of Waterloo, Waterloo, Ontario N2L 3G1, Canada\label{aff78}
\and
Department of Physics and Astronomy, University of Waterloo, Waterloo, Ontario N2L 3G1, Canada\label{aff79}
\and
Perimeter Institute for Theoretical Physics, Waterloo, Ontario N2L 2Y5, Canada\label{aff80}
\and
Space Science Data Center, Italian Space Agency, via del Politecnico snc, 00133 Roma, Italy\label{aff81}
\and
Institute of Space Science, Str. Atomistilor, nr. 409 M\u{a}gurele, Ilfov, 077125, Romania\label{aff82}
\and
Dipartimento di Fisica e Astronomia "G. Galilei", Universit\`a di Padova, Via Marzolo 8, 35131 Padova, Italy\label{aff83}
\and
INFN-Padova, Via Marzolo 8, 35131 Padova, Italy\label{aff84}
\and
Instituto de F\'isica Te\'orica UAM-CSIC, Campus de Cantoblanco, 28049 Madrid, Spain\label{aff85}
\and
Institut de Recherche en Astrophysique et Plan\'etologie (IRAP), Universit\'e de Toulouse, CNRS, UPS, CNES, 14 Av. Edouard Belin, 31400 Toulouse, France\label{aff86}
\and
Universit\'e St Joseph; Faculty of Sciences, Beirut, Lebanon\label{aff87}
\and
Departamento de F\'isica, FCFM, Universidad de Chile, Blanco Encalada 2008, Santiago, Chile\label{aff88}
\and
Universit\"at Innsbruck, Institut f\"ur Astro- und Teilchenphysik, Technikerstr. 25/8, 6020 Innsbruck, Austria\label{aff89}
\and
Department of Physics and Helsinki Institute of Physics, Gustaf H\"allstr\"omin katu 2, University of Helsinki, 00014 Helsinki, Finland\label{aff90}
\and
Department of Physics, Royal Holloway, University of London, Surrey TW20 0EX, UK\label{aff91}
\and
Departamento de F\'isica, Faculdade de Ci\^encias, Universidade de Lisboa, Edif\'icio C8, Campo Grande, PT1749-016 Lisboa, Portugal\label{aff92}
\and
Instituto de Astrof\'isica e Ci\^encias do Espa\c{c}o, Faculdade de Ci\^encias, Universidade de Lisboa, Tapada da Ajuda, 1349-018 Lisboa, Portugal\label{aff93}
\and
Cosmic Dawn Center (DAWN)\label{aff94}
\and
Niels Bohr Institute, University of Copenhagen, Jagtvej 128, 2200 Copenhagen, Denmark\label{aff95}
\and
Universidad Polit\'ecnica de Cartagena, Departamento de Electr\'onica y Tecnolog\'ia de Computadoras,  Plaza del Hospital 1, 30202 Cartagena, Spain\label{aff96}
\and
European University of Technology EUt+, European Union\label{aff97}
\and
Institute of Space Sciences (ICE, CSIC), Campus UAB, Carrer de Can Magrans, s/n, 08193 Barcelona, Spain\label{aff98}
\and
Institut d'Estudis Espacials de Catalunya (IEEC),  Edifici RDIT, Campus UPC, 08860 Castelldefels, Barcelona, Spain\label{aff99}
\and
INFN-Bologna, Via Irnerio 46, 40126 Bologna, Italy\label{aff100}
\and
Caltech/IPAC, 1200 E. California Blvd., Pasadena, CA 91125, USA\label{aff101}
\and
Dipartimento di Fisica - Sezione di Astronomia, Universit\`a di Trieste, Via Tiepolo 11, 34131 Trieste, Italy\label{aff102}
\and
Oskar Klein Centre for Cosmoparticle Physics, Department of Physics, Stockholm University, Stockholm, SE-106 91, Sweden\label{aff103}
\and
Astrophysics Group, Blackett Laboratory, Imperial College London, London SW7 2AZ, UK\label{aff104}}    